\documentstyle{mn}

\newcommand{\etal}{{\em et al.}}                      % puts et al. in italics.

\newcommand{\be}{\vspace{0mm}\begin{equation}}
\newcommand{\ee}{\end{equation}\vspace{0mm}}
\newcommand{\bs}{\vspace{0mm}\begin{eqsub}}
\newcommand{\es}{\end{eqsub}\vspace{0mm}}
\newcommand{\bbs}{\vspace{0mm}\refstepcounter{equation} \begin{eqsub}}

\begin{document}

\title{High-redshift radio galaxies and quasars at sub-millimetre
wavelengths: assessing their evolutionary status.}

\author[D.H. Hughes, J.S. Dunlop, S. Rawlings]
       {D. H. Hughes$^{1,2}$, J. S. Dunlop$^{1}$ \& S. Rawlings$^{2}$ \\
       $1$.\ Institute for Astronomy, Dept. of Physics \& Astronomy,
       University of Edinburgh, Royal Observatory, Edinburgh, EH9 3HJ, U.K.\\
       $2$.\ Astrophysics, Nuclear Physics Laboratory, Oxford University,
       Keble Rd., Oxford OX1 3RH, U.K.}

\date{ }

\maketitle

\begin{abstract}
\noindent 
We present new results of a study of the sub-millimetre continuum 
emission from a sample of 9 radio galaxies and 4 radio-quiet quasars 
at redshifts $z = 0.75 - 4.26$. The observations were made at
800$\mu m$, using the 
single-element bolometer UKT14 on the James Clerk Maxwell Telescope (JCMT), 
reaching a typical r.m.s. sensitivity of $\sigma_{rms} \sim 4$\,mJy
and represent some of the deepest submillimetre extragalactic
measurements made to date. 
Three detections were achieved, of which two are secure 
(4C41.17, Dunlop {\it et al.} (1994) and H1413$+$117) and one 
(53W002) is tentative, whilst comparable
upper-limits were obtained for 7 of the 10 remaining sources.  
We use these data as the motivation for a detailed discussion of
the conversion from submillimetre
and millimetre continuum fluxes to dust/gas masses and star formation rates at 
high-redshift, 
and determine these quantities from our own and other data on 
high-redshift radio-galaxies and quasars. 
In particular we have investigated the impact of
the four main sources of uncertainty in deriving physical quantities from
such data, namely i) potential 
contamination by galactic cirrus, ii) uncertainty in the value of 
the dust rest-frequency mass-absorption
coefficient, iii) difficulty in constraining the dust temperature, and
iv) estimation of the appropriate gas:dust ratio in high-redshift
objects. Our discussion emphasises how important it will be to quantify
and, where possible, minimize such uncertainties (via, for example, appropriate
observational strategies) in order to fully capitalize on the ten-fold
improvement in sensitivity offered by the imminent arrival of the 
next generation of bolometer arrays, such as SCUBA 
on the JCMT. 

Taking these uncertainties into account we show that whilst the
high-redshift galaxies discussed in this paper are amongst the most
dust-rich and luminous objects discovered to date, 
their far-infrared  properties are more comparable with those of
the most luminous nearby 
interacting galaxies than with those expected of prim\ae val
giant ellipticals. This conclusion is rather insensitive to the
adopted dust temperature, and the appropriateness of our adopted gas:dust ratio 
is supported by the molecular line detections of lensed objects.
Indeed, despite all of the uncertainties peculiar to studying galaxy
evolution at sub-millimetre wavelengths, the current
uncertainty over the true value of 
$\Omega_0$ is probably the most important factor affecting our conclusions.

\vspace*{0.2in}


\end{abstract}

\section{Introduction}

Over the past four years, we have undertaken a number of
sub-millimetre observations 
of high-redshift radio galaxies and quasars (using the
single channel 
$^3$He bolometer UKT14 on the JCMT) aimed at determining the evolutionary
status of these objects on the basis of the strength of their
rest-frame far-infrared continuum emission. Since the arrival of the 
new generation of sub-millimetre bolometer arrays is now imminent, this
seems an appropriate time to summarize 
the existing observational results (including our own data presented
here), to review the main sources of
uncertainty which afflict the meaningful physical interpretation of such
data, and, in the light of recent Keck and HST results (Steidel \etal
1996, Illingworth 1997), 
to re-appraise the importance of deep sub-millimetre
observations for the study of galaxy evolution and formation (Hughes 1996).
  
The original motivation for attempting to detect sub-millimetre emission 
in high-redshift galaxies, despite the relative insensitivity 
of existing sub-millimetre detectors 
was the realistic possibility that at least some massive 
galaxies formed the majority of their stars in a relatively short ($< 1$
Gyr) starburst at high redshift, and that the resulting necessarily high
star-formation rates ($\simeq 100 \rightarrow 1000 M_{\odot} yr^{-1}$)
could give rise to strong rest-frame far-infrared thermal emission from
dust.

This motivation remains as strong as ever despite the
recent discovery at {\it optical} wavelengths of a 
population of star-forming galaxies at redshifts $z \sim 3$ by 
Steidel {\it et al.} (1996). While the space-density of these Lyman-limit
galaxies at $z \simeq 3$ indicates that we may be seeing the progenitors
of a substantial fraction of
present-day bright galaxies, their rather moderate star formation rates
($\sim 10 M_{\odot} yr^{-1}$) 
would have to be sustained for virtually a Hubble time to
produce the most massive elliptical galaxies which exist in the
present-day Universe (of stellar mass $\simeq 10^{12} M_{\odot}$).
The implication is that, while in the Lyman-limit galaxies we may be
seeing the formation of the bulges of spiral galaxies and the cores of
some ellipticals (an interpretation supported by their compact 
spheroidal appearance; 
Giavalisco {\it et al.} 1996), certainly for the most massive galaxies the
star-formation rates must have been substantially greater 
($\ge 100 M_{\odot} yr^{-1}$) at higher redshifts. 

Indeed, we now possess rather strong evidence that this must have been
the case. 
First, the basic properties of present-day elliptical galaxies, namely
low molecular gas and dust masses, $< 10^{8} M_{\odot}$ 
(Lees {\it et al.} 1991, Knapp \& Patten 1991, Wilkind \& Henkel 1995),
enormous stellar masses $\sim 10^{11} - 10^{12} M_{\odot}$, determined from
the K-band luminosities ($\gg 2 \times L^{\ast}$, Taylor {\it et al.}
1996) of the host galaxies of low-z radio galaxies, 
RLQs and the most luminous RQQs,
and uniform optical-IR colours (Bower, Lucey \& Ellis 1992) 
that are dominated by a well-evolved stellar
population, indicate that the bulk of their stars were formed in a
relatively short-lived star-burst at high redshift.
Second, rather more severe constraints on what the phrase `high-redshift'
must actually mean have been recently provided by Dunlop {\it et al.} (1996),
who have presented spectroscopic evidence that star formation
ceased in a $z=1.55$ radio galaxy at least 3.5 Gyr prior to the epoch at
which the galaxy is observed. Such a large age 
formally excludes an Einstein--de-Sitter cosmology for a 
Hubble constant $H_{0} > 50 ~ \rm km s^{-1}
Mpc^{-1}$ and thus indicates 
an extreme formation redshift ($z \stackrel{>}{_\sim} 5$) for this galaxy
in almost any cosmological model. 
This object is not a freak; a second, apparently even older radio galaxy 
has now been discovered at a similar redshift (Dunlop 1997) and
spectroscopy of radio quiet ellipticals in the cluster around the $z =
1.206$ radio
galaxy 3C324 indicate similarly old ages for the vast majority of their
stellar populations (Dickinson 1997).
Third, a radio galaxy has recently been discovered at $z = 4.41$ 
(Rawlings {\it et al.} 1996); this demonstrates that at least some massive galaxies 
were in place at early epochs, and its properties ({\it e.g.} low
rest-frame ultraviolet flux) are not obviously those of a young star-forming
system. In summary, despite the important discovery of the star-forming 
population at $z \simeq 3$ (Steidel \etal 1996), a growing body of evidence in fact 
indicates that we have yet to
discover the formation epoch of the most massive galaxies and that, given
the available cosmological time, the formation of these objects must have
been shortlived and hence should be spectacular in at least one
wavelength regime.

These last two studies illustrate the key r\^{o}le which radio galaxies
continue to play in studies of high-redshift galaxy evolution and
formation. The main reason for this is that all low-redshift 
radio galaxies are associated with giant elliptical galaxies 
(Owen \& Laing 1989), and thus it is reasonable to assume that they 
can be used to trace the evolution of the most massive galaxies back to
high redshifts, and early cosmic epochs. 

However, while we can be reasonably confident that the hosts of high-redshift
radio sources are the progenitors of giant elliptical galaxies, recent years 
have seen a growing concern that, at least in the most powerful sources,
the direct or indirect effects of their active nuclei might stymie any 
attempt to determine their evolutionary status at optical-infrared wavelengths.
For example, the discovery of multi-modal optical structures aligned along the radio
axis of high-redshift radio galaxies (Chambers {\it et al.} 1987;
McCarthy {\it et al.} 1987) seemed at first sight to indicate 
the assembly of giant elliptical galaxies from the
merging of lower-mass clumps at relatively recent redshifts, 
as expected in certain hierarchical models of structure formation. 
However, in subsequent years there has been 
much debate about whether this is true, and about the evolutionary status of 
high-redshift radio galaxies in general. Early arguments that $z > 2$ radio
galaxies contained old evolved stars ({\it e.g.} Lilly 1988; Lilly 1989) have been
questioned ({\it e.g} Chambers \& Charlot 1990; Eales \& Rawlings 1993) but
there has been a distinct lack of proof that any of the stellar systems
are demonstrably young (although arguments have been made that this is the 
case; {\it e.g.} 53W002 -- Windhorst {\it et al.} 1991; 0902$+$34 -- 
Eales {\it et al.} 1993). The few hints of `prim\ae vality', 
gleaned from ultraviolet--infrared observations, have remained hard to
prove largely because of the way in which high-redshift radio 
galaxies are typically selected by virtue of their extreme radio luminosity.
An inescapable consequence of high radio luminosity appears to
be significant non-stellar emission across the ultraviolet, optical
and infrared bands ({\it e.g.} Rawlings \& Saunders 1991; Dunlop \& Peacock
1993; Eales \& Rawlings 1996). In other words huge narrow emission lines,
large ultraviolet luminosities and multi-modal structures 
(which naively one might interpret as evidence for extremely high 
star-formation rates
and dynamically young systems) are now often attributed to 
to the action of the powerful jets or quasar light emanating from the
active nuclei of these extreme objects.
On the other hand, even when radio galaxies are 
faint ({\it e.g.} the $z=4.41$ radio galaxy studied by Rawlings \etal\ 1996)
it is impossible to disprove huge star-formation rates since, as is the
case for some local ultra-luminous IRAS galaxies (Sanders \etal\ 1988), 
star formation activity may be enshrouded in dust.

This latter point provides the prime motivation for the study described here.
A more robust signature of massive star formation than 
ultraviolet continuum and emission lines is intense FIR
emission from dusty, molecular material, where the rate of dust production 
is proportional to the star formation rate. The dust is heated primarily 
by the embedded  O and B stars which evolve quickly and disperse their 
surrounding material on similarly short timescales 
($\sim 10^{7}$yrs., Wang 1991). Hence the FIR luminosity provides 
a measure of the current formation rate (SFR) of massive stars,

\be SFR = \Psi 10^{-10} \frac{L_{FIR}}{L_{\odot}} \hspace{5mm}  
M_{\odot} yr^{-1} \ee

\noindent where $\Psi = 0.8 - 2.1$ (Scoville \& Young 1983, Thronson \& 
Telesco 1986). 
If galaxies at high redshift have FIR luminosities comparable to
or greater than low-z ULIRGS ($L_{FIR}/L_{\odot} \ge 10^{12}$),
hence SFRs $> 100 M_{\odot} yr^{-1}$, then it is possible that they convert 
$10^{11} -10^{12} M_{\odot}$ of gas into stars in a burst of
duration  $< 1$\,Gyr.

In the Milky Way and local disc galaxies a significant fraction ($30\%$) of the
bolometric luminosity ($L_{bol}$) 
is re-radiated at FIR wavelengths, and hence the 
SFR (Miller \& Scalo 1979, Kennicutt 1983) and the ratio
$L_{FIR}/L_{bol}$ cannot have evolved much with look-back time.
However the situation is very different for elliptical galaxies.
Mazzei, de Zotti \& Xu (1994) have 
modelled the photometric evolution of elliptical galaxies and show that, 
whilst at the current epoch ellipticals emit $<1\%$ of their 
bolometric luminosity at FIR wavelengths, within the first $1-2$\,Gyr of
their formation this fraction was significantly higher with 
$L_{FIR}/L_{bol} \sim 0.3$ (see figure 1).

Whilst the details of the
evolution are sensitive to the assumed initial mass function (IMF) and SFR, 
(where the steeper IMF and 
higher SFR produces a more luminous, but shorter, burst of starformation),
it is hard to escape the conclusion
that the formation of giant elliptical galaxies is expected to 
be a  spectacular and luminous phenomenon in the rest-frame far-infrared.
Indeed this should be true irrespective of 
whether ellipticals form via the collapse of a single gaseous halo, or
grow through the rapid merging of smaller mass clumps at high-z, particularly
since the mass dependence of metallicity and mass:light ratio in
elliptical galaxies indicates that early star-formation in massive
ellipticals should be strongly biassed towards high-mass stars (Zepf \&
Silk 1996).

\begin{figure}
\begin{picture}(220,215)
\put(0,0){\includegraphics{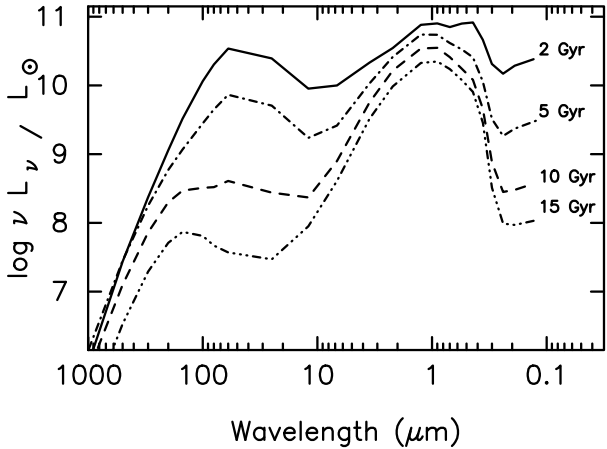}}
\end{picture}
\caption{
The anticipated evolution of the UV to millimetre wavelength 
rest-frame spectral energy 
distribution of elliptical galaxies. Models corresponding to
galactic ages between 2 - 15 Gyr are shown, assuming a Salpeter IMF
with a lower mass limit, $m_{l} = 0.01 M_{\odot}$. The SFR 
($\psi(t) = (m_{gas}/m_{gal}) \psi_{0}$)
is proportional to the fractional mass of gas in the galaxy
and assumes an intial SFR $\psi_{0} = 100 M_{\odot}$/yr.    
The predicted SEDs are taken from Mazzei, de Zotti \& Xu (1994).}
\end{figure}

\begin{figure*}
\begin{picture}(200,305)
\put(-120,0){\includegraphics{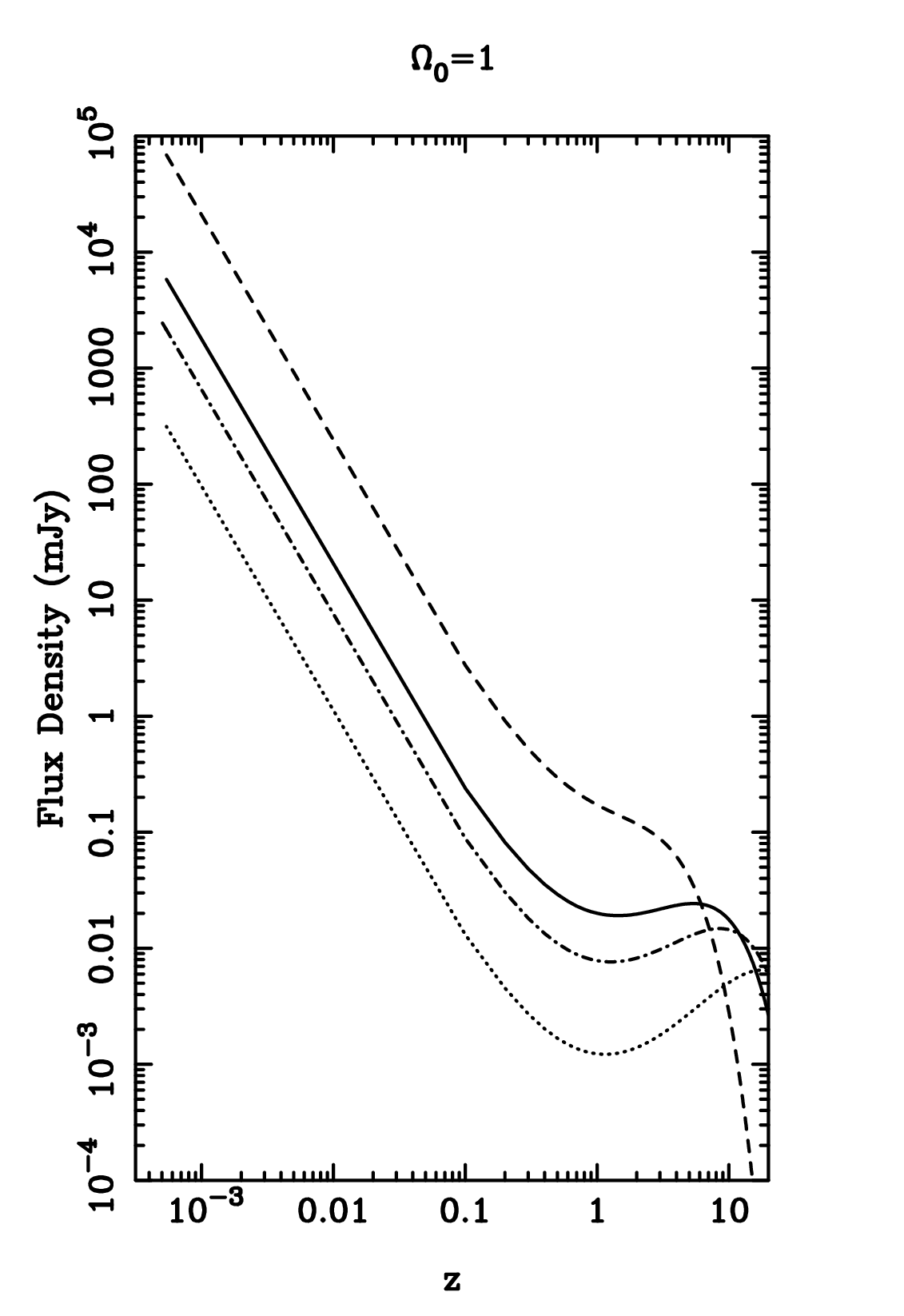}}
\put(140,0){\includegraphics{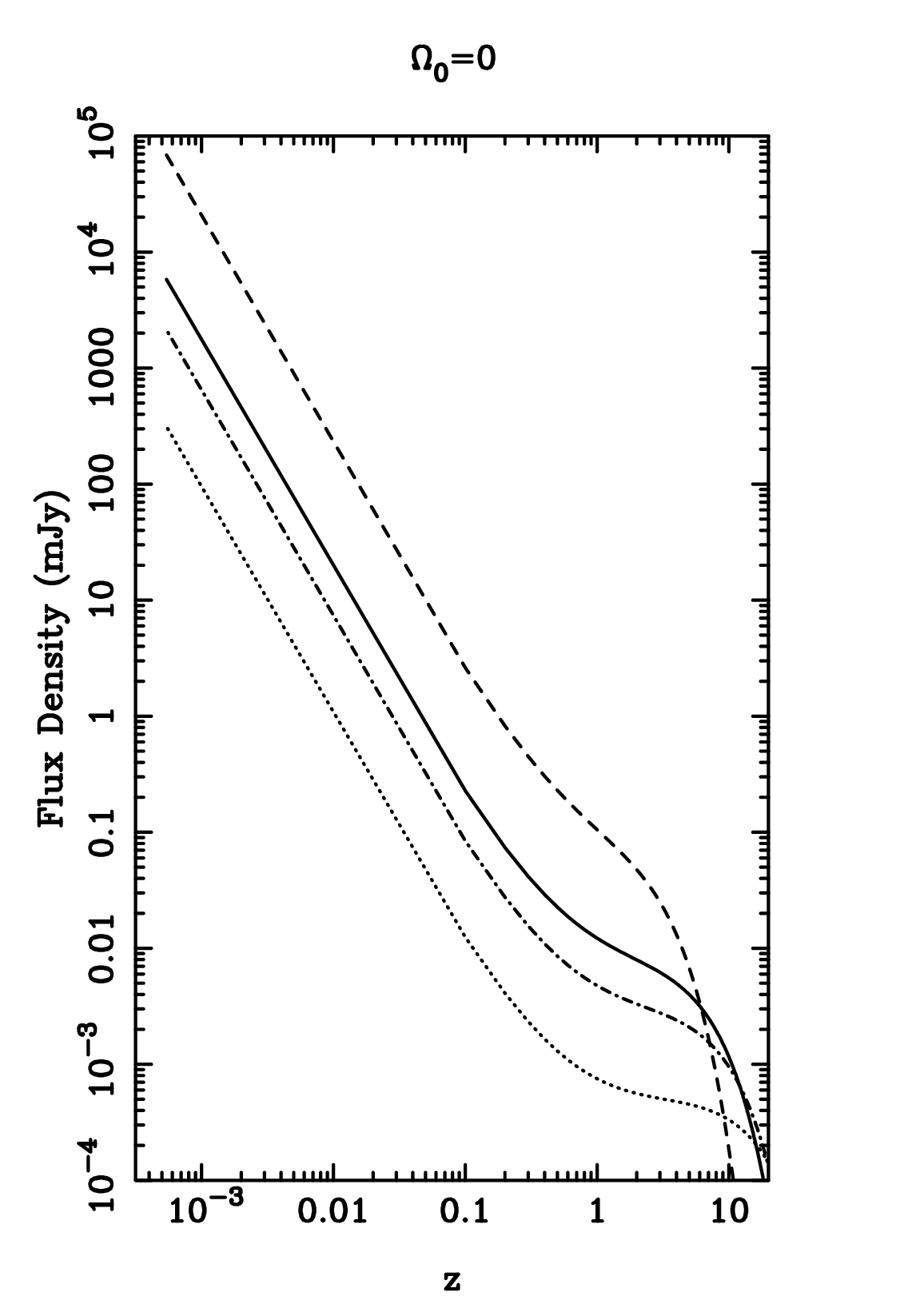}}
\end{picture}
\caption{Dependence of the observed flux density on redshift 
of the starburst galaxy M82 (Hughes {\it et al. 1994}) 
at 350$\mu m$ (dashed line), 800$\mu m$ (solid line), 1100$\mu m$ 
(dashed-dotted line) and 2000$\mu m$ (dotted line) for $\Omega_{0} = 1$
(left-hand plot)
and $\Omega_{0} = 0$ (right-hand plot) 
assuming $H_0 = 50$ km s$^{-1}$ Mpc$^{-1}$.}
\end{figure*}

Any attempt to detect this thermal radiation from 
dust in high-redshift radio galaxies would fail
were it not for the fact that the expected far-infrared emission peak
(at $\lambda \sim 60-100\mu m$), 
due to grains radiating at temperatures of 30--70\,K, is shifted into
the submillimetre region ($\lambda > 300\mu m$) at the redshifts of 
interest. Figure 2a demonstrates that, in an Einstein-de Sitter universe 
and at wavelengths 
$\ge 800\mu m$, the effects of cosmological dimming on a typical starburst
galaxy spectrum are offset between redshifts $z = 1 \rightarrow 10$ 
by the strongly negative k-correction (which arises as the submillimetre
filters effectively climb 
the Rayleigh-Jeans tail of the thermal dust emission with increasing redshift).
The situation for a low density universe is not as advantageous (figure\,2b), 
where the observed
flux density for a redshifted object continues to fall, albeit more
slowly, in all submillimetre and millimetre wavelength passbands.

In this paper we report on an attempt to exploit the `detectability' 
of thermal emission from dust at high redshift to
determine the evolutionary status of known high-redshift radio galaxies
and quasars. The aim is to use sub-millimetre
continuum photometry to estimate or at least constrain the mass of dust
in high-redshift objects, with a view to determining both the star
formation rate and, more importantly, the mass of gas which has yet to be
turned into stars in these potentially young galaxies.

While gas masses can in principle be derived from molecular line
observations, we have chosen to concentrate on a sub-millimetre continuum 
approach for three reasons.
First, comparable assumptions and uncertainties also complicate attempts
to derive an accurate $H_2$ mass from detections of CO emission. Second,
despite determined efforts, with the exception of BR1202$-$0725 (Omont \etal\
1996, Ohta \etal\ 1996), no significant detection of molecular 
gas (through, {\it e.g.} CO  line transitions) has been achieved in any high-z 
radio-galaxy or quasar (Evans {\it et al.} 1996, van Ojik {\it et
al.} 1997, Barvainis \& Antonucci 1996) unless it has been greatly
magnified by gravitational lensing (Barvainis \etal\ 1994,
Eisenhardt \etal\ 1996). Third it is in
sub-millimetre {\it continuum} astronomy that the greatest advances in
sensitivity are likely to be forthcoming in the immediate future.

As explained above, we have chosen to 
concentrate on radio galaxies because of their likely association
with the high-redshift counterparts of massive elliptical galaxies,
but we have also observed a few radio-quiet quasars in an attempt to
confirm previously published detections. 
Our sub-millimetre observations are described in Section 2. 
In section 3, we present a detailed point-by-point
analysis of the various uncertainties which afflict attempts to
accurately estimate the dust mass of a high-redshift galaxy from
sub-millimetre observations. Then in section 4 we derive our best
estimates of the dust masses of these high-redshift objects and use this
information to assess their evolutionary status, by estimation of both
the star formation rate, and the amount of molecular gas which, at the
time of observation, has yet to be converted into stars. Finally, in
section 5 we discuss
explicitly, and endeavour to quantify the realistic uncertainties in
these extrapolations, in particular the gas:dust ratio at high-redshift.

Throughout the paper we assume $q_{0} = 0.5$
and $H_{0} = 50$kms$^{-1}$\,Mpc$^{-1}$  unless stated otherwise,  
and correct previously published  
data to the same cosmology to allow a comparison of all physical quantities.

\section{Submillimetre Observations}

\subsection{Target selection strategy}

As figure 2 illustrates, if one wishes to attempt to detect high-redshift
($z > 1$) objects at sub-millimetre wavelengths, little if anything is lost 
in terms of sensitivity between $z \simeq 1$ and $z \simeq 10$.
Consequently, since the star-formation rate in elliptical galaxies is
expected to be highest at early times (figure 1) it makes sense to start
with the most distant known objects. We thus selected our primary targets
(see Table\,1) 
from among the most distant known radio galaxies, concentrating on
objects which have been well-studied in the optical-infrared while also
attempting to span a meaningful range of radio luminosities and redshifts. 
4C41.17 and 8C1435$+$643 are extremely luminous radio sources and each, 
at the time they were observed,
held the title of the most distant
known galaxy in the Universe. B20902$+$34 is an extremely well-studied 
object and has been hailed as one
of the best known candidates for a primeval galaxy (Eales {\it et al.}
1993). 6C0032$+$412 and 
MG2141$+$192 are two of the most distant known galaxies at more moderate
radio luminosities, while 53W002 is the most distant mJy radio galaxy
(Windhorst \etal\ 1991). 
3C257 was selected because it is the most distant 3CR radio 
galaxy and because near-infrared spectroscopy has indicated that 
Lyman-$\alpha$ emission in this source is attenuated by dust. Two
additional radio galaxies at intermediate redshifts ($z \simeq 1$) 
were selected because of circumstantial evidence suggesting the presence
of dust; 3C318 is the most distant radio galaxy detected by IRAS
(Heckman \etal 1994) and 3C65 has an extremely red r--K colour (Lilly
1989, Dunlop \& Peacock 1993).
Lastly we observed 4 RQQs, again spanning the redshift
range ($z = 1 \rightarrow 4$), both to attempt to confirm or refute
claimed IRAM detections at millimetre wavelengths, and to enable us to
at least make a preliminary attempt to compare the far-infrared
properties of radio-loud and radio-quiet objects at high redshift.
  
\begin{table*}
\caption{800 $\mu m$ photometry of high-redshift radio galaxies (RGs)
and radio-quiet quasars (RQQs). The flux density limits quoted in column
6 are given at the 3$\sigma$ level. The references in column\,7 give
the positions, redshifts and details of other relevant observations.}  
\begin{tabular}{llccccl}
\hline
Source name    & type & \,\,\,$z$\,\,\,     &  R.A. (J2000) &  Dec (J2000)   & $S_{800\mu m}$ (mJy)        & refs \\ \hline
               &     &          &             &                &                             &      \\
3C 318         & RG  & 0.752    & 15 20 05.49 & +20 16 05.1    &  $<$33  & 1\\
3C 65          & RG  & 1.176    & 02 23 43.48 & +40 00 52.7    &  $<$11              & 1,2\\
PG1634+706     & RQQ & 1.334    & 16 34 28.97 & +70 31 32.4    &  $<$47 & 3 \\
53W002         & RG  & 2.390    & 17 14 14.78 & +50 15 30.4    & 6.9 $\pm$ 2.3 & 4,5\\
3C 257         & RG  & 2.474    & 11 23 09.40 & +05 30 17.8    &  $<$11              & 6,7,8\\
H1413+117     &  RQQ & 2.546    & 14 15 46.26 & +11 29 43.7    &  66 $\pm$ 7   &  9     \\
2132$+$0126    & RQQ & 3.194    & 21 35 10.61 & +01 39 31.3    &  $<$12              & 10\\ 
B2 0902$+$34   & RG  & 3.391    & 09 05 30.10 & +34 07 57.3    &  $<$14  & 6,11,12\\
MG 2141$+$192  & RG  & 3.594    & 21 44 07.52 & +19 29 14.2    &  $<$11              &  6,8\\
0345$+$0130    & RQQ & 3.638    & 03 48 02.29 & +01 39 18.4    &  $<$25              & 10 \\
6C 0032$+$412  & RG  & 3.665    & 00 34 53.09 & +41 31 31.5    &  $<$14  & 8,13\\
4C 41.17       & RG  & 3.800    & 06 50 52.36 & +41 30 31.5    & 17.4$\pm$ 3.1 & 14,15,6\\
8C1435+643     & RG  & 4.252    & 14 36 37.19 & +63 19 14.2    &  $<$13             & 16,17\\
               &     &          &             &                &                             & \\ \hline
\end{tabular}

$1.$ Laing {\it et al.} (1983)
$2.$ Stockton {\it et al.} (1995)
$3.$ Schmidt \& Green  (1983)
$4.$ Windhorst {\it et al.} (1991),
$5.$ Windhorst {\it et al.} (1994),
$6.$ Eales \& Rawlings (1993),
$7.$ McCarthy \etal (1995),
$8.$ Eales \& Rawlings (1996),
$9.$ Magain \etal\ (1988) 
$10.$ Schneider {\it et al.} (1991)
$11.$ Lilly (1988),
$12.$ Eisenhardt \& Dickinson (1992),
$13.$ Rawlings \etal\  in prep,
$14.$ Chambers {\it et al.} (1990),
$15.$ Miley {\it et al.} (1992).
$16.$ Lacy \etal\ (1994)
$17.$ Spinrad {\it et al.} (1995)

\end{table*}

\subsection{Selection of observing wavelength}

Obviously we have no {\em a priori} knowledge of the exact spectral
shape of the continuum at FIR-mm wavelengths in the rest frame of the 
high-redshift radio galaxies. However, 
since our aim is to test whether these galaxies 
are undergoing an intense burst of starformation at early epochs, 
it seems reasonable to assume that
their rest-frame FIR-mm continuum spectra will be similar in form 
to those of massive galactic starforming regions  and low-z starburst
galaxies. At the highest redshifts the FIR spectral peaks of such 
starforming regions fall conveniently into the region of 
two submillimetre wavelength atmospheric windows at 
350$\mu m$ and 450$\mu m$ and it might seem that observations would be 
most profitably made at these wavelengths. However the 750$\mu m$ and 
850$\mu m$ atmospheric windows provide more suitable and, in practice, 
preferential alternatives. At redshifts $z > 2$ the observed 
flux density ratio 
$S_{400\mu m}$/$S_{800\mu m}$ of a starburst galaxy spectrum is significantly 
smaller than that at $z \sim 0$ (Figure 3) and is no longer 
sufficient to offset the observational disadvantages of increased sky
noise and reduced sky transparency, $A^{trans}$, at 
450$\mu m$ compared to 800$\mu m$, under even the dryest atmospheric
conditions ($p.w.v. \simeq 0.5$\,mm) when $A^{trans}(450\mu m/800\mu m)$
is $\le 0.7$.  

\begin{figure}
\begin{picture}(220,230)
\put(0,0){\includegraphics{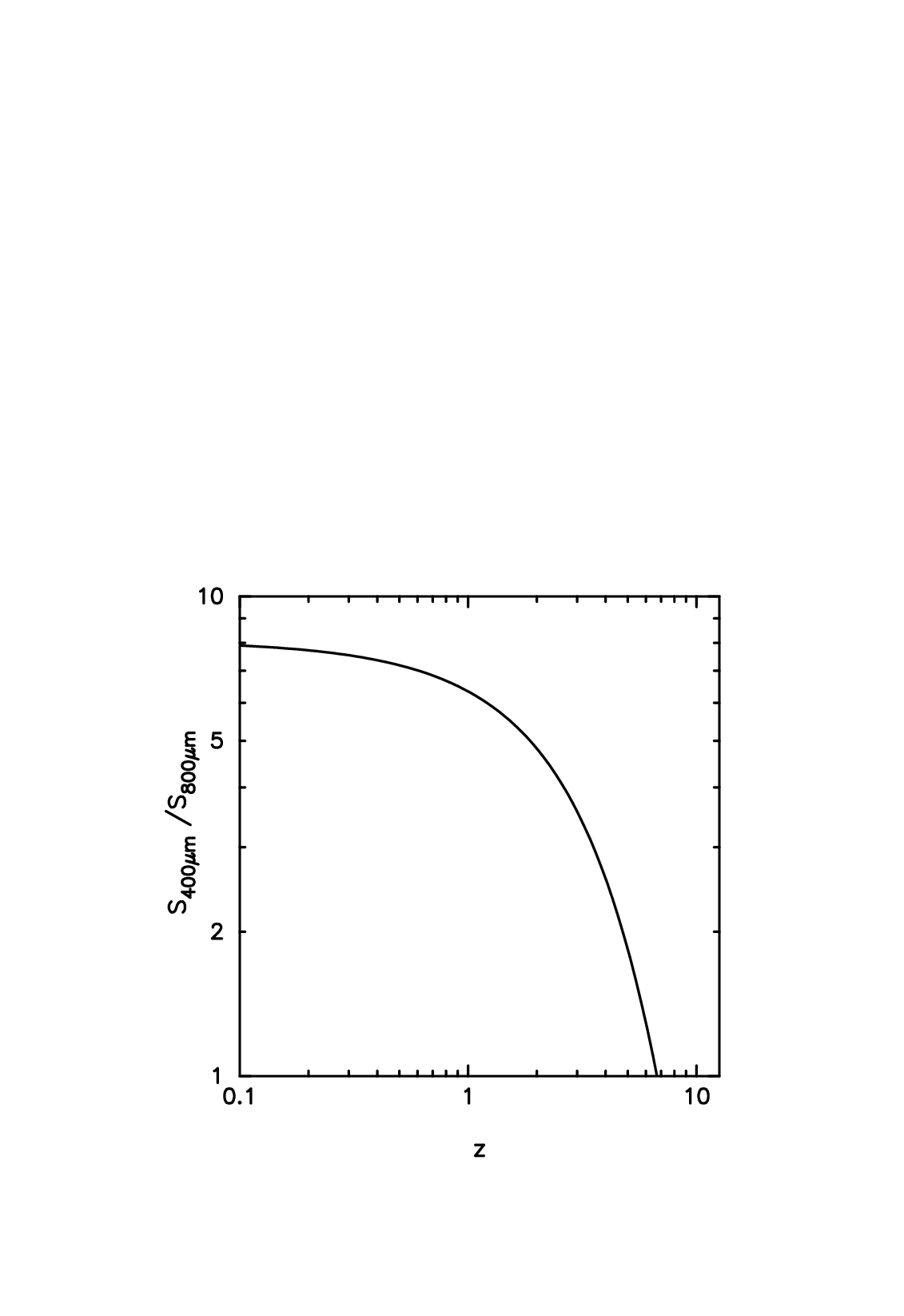}}
\end{picture}
\caption{The predicted redshift 
dependence of the observed flux-density ratio $S_{400\mu m}/S_{800\mu m}$
calculated from the isothermal dust model fitted to the FIR-mm continuum 
data of M82 (Hughes {\it et al.} 1994).}
\end{figure}

Therefore all observations were initally made
in the 850$\mu m$ and 750$\mu m$ windows using a broad-band filter
at $\simeq 800\mu m$
($\lambda_{c}= 781\mu m$, $\Delta \lambda \sim 200\mu m$). In the event
of a detection at 800$\mu m$ follow-up observations were made at
450$\mu m$.
An additional theoretical point also strongly favours the use of a filter at 
$\sim 800\mu m$ over shorter wavelength filters 
since, as discussed in \S 3.3.2, a single photometric measurement
at $\lambda \ll 800\mu m$ is  
virtually useless for constraining the dust masses in sources at $z > 1$.

\subsection{Observations and Results}

Sensitive continuum observations of radio galaxies and RQQs  
in the redshift range $0.8 \leq z \leq 4.3$
were made at 800$\mu m$ using the single-element $^{3}$He 
bolometer UKT14 (Duncan {\it et al.} 1990) on
the JCMT during a series of runs between April 1993
and February 1996. Data was taken only under
excellent submillimetre observing conditions
($\tau_{800\mu m} \simeq 0.35$).
A 65\,mm focal-plane aperture was used for all observations and the resulting
beamsize of 16.5 arcsec corresponds to  physical 
diameters of $\simeq 140 \rightarrow 100$\,kpc at $z = 1 \rightarrow 4$.
An azimuthal chop throw of 60 arcsec at a frequency of 7.81\,Hz
was used to subtract the sky emission. 
For each source a number of repeated, short (10--20 minute) observations were 
concatenated using a method described by Hughes {\it et al.} (1993).
Total on-source integration times were typically 3--5 hours.
Primary calibration was performed against Uranus and Mars, with secondary
calibration against a variety of AGB stars and compact HII regions 
(Sandell 1994). 
The data collected to date with UKT14 are summarized in Table\,1.
Note that, after the successful commissioning of the 0.1K bolometer array
SCUBA ($> 1997$), one can expect typical $3\sigma$ 
sensitivities of 12\,mJy and 1\,mJy at 450$\mu m$ and 800$\mu m$ respectively
with an on-source integration of 3\,hours.

\subsubsection{High-redshift radio galaxies}

A highlight of this study has been the clear detection of 4C41.17 at 
$800\mu m$, together
with a significant upper limit of $3\sigma < 56$mJy at 450$\mu m$ (Dunlop
{\it et al.} 1994). This detection was subsequently confirmed by an
IRAM observation at 1.25mm (Chini \& Kr\"{u}gel 1994) at a level consistent
with the thermal dust model presented by Dunlop {\it et al.} (1994).
We have also achieved a marginal $3\sigma$ detection at 800$\mu m$ of 
53W002 at a considerably fainter flux density of 
$6.9 \pm 2.3$\,mJy. 
Yamada {\it et al.} (1995) have a similarly tentative detection of
$^{12}$CO(1--0) in 53W002 leading to an H$_{2}$ gas mass of $2 \times 
10^{12}$\,M$_{\odot}$. 
Whilst we have had no opportunity to confirm our continuum 
result, treating the $800\mu m$ observation as a limit of $3\sigma \leq 7$\,mJy
($M_{d} = 1.5 \times 10^{8} M_{\odot}$, see \S 3.2)
allows us to place an unlikely lower limit on the gas-to-dust ratio of
$> 13000$ and hence casts doubt on the validity of the CO detection.
At the time of observation it seemed somewhat 
surprising that no 800$\mu m$ detection of B20902+34
was obtained given the claimed IRAM detection at 1.3\,mm (Chini \& Kr\"{u}gel
1996). However recent 105\,GHz observations (Yun \& Scoville 1996, 
Downes \etal\ 1996) now suggest that a significant fraction (probably the
majority) of the
1.3\,mm continuum in B20902$+$34 is due to non-thermal emission (see \S 3.1).
No detections were achieved for the remaining six radio galaxies at z$ >
2$; our non-detection of 8C1435$+$643 could be regarded as casting some
doubt on the validity of the IRAM detection by Ivison (1995), but in fact
our 3 sigma upper limit for this source is still just consistent with a
grey-body fitted through the 1.25mm data point (see section 3.1).

\subsubsection{High redshift radio-quiet quasars.}

Continuum  observations at 800$\mu m$ (Barvainis {\it et al.}
1992;  Isaak {\it et al.} 1994)
and at 1.25\,mm (Andreani, La Franca \& Cristiani 1993, McMahon {\it et al.} 
1994, Ivison 1995, Omont {\it et al.} 1996)
have suggested that thermal emission from dust has been detected in 
high-redshift radio-quiet quasars. 
In an attempt to confirm some of these more marginal results and facilitate 
direct comparison with our sample of high-redshift radio galaxies,
we made 800$\mu m$ observations of two 
(2132$+$0126 and 0345$+$0130) of the three radio-quiet quasars 
detected by Andreani {\it et al.} (1993).
In 2132$+$0126 the 1.25\,mm detection (11.7 mJy) and 
our measured 800$\mu m$ flux density limit ($< 12$mJy) produce a
spectral index $\alpha <0.06$ ($S_{\nu} \propto \nu^{\alpha}$)
over the rest-frame wavelength range $298\mu m \rightarrow 191\mu m$.
This is inconsistent with that expected from 
a thermal dust spectrum with a grain temperature $T \ge 20$\,K.
Therefore our result casts serious doubt either on the 1.25\,mm IRAM 
detections 
or on the proposed thermal nature of the sub-millimetre 
continuum, unless the dust grains are significantly colder than observed 
in low-redshift quasars (Chini \etal 1989, Hughes \etal 1993).
Our flux limit of $3\sigma < 25$\,mJy at 800$\mu m$ for 0345$+$0130
is just consistent with the flux density expected from an extrapolation
of the 1.25\,mm detection (assuming an isothermal 50\,K spectrum with a grain
emissivity index $\beta =2$) 
and thus provides no additional
constraint on the nature of the FIR spectrum
or the dust mass in this object.

In the light of the recent detections of the {\em Cloverleaf} quasar
(H1413$+$117) at 1.25mm, 100$\mu m$ and 60$\mu m$ (Barvainis {\it et al.}
1995), which indicated it to have a peculiar SED (figure 4), we dearchived
and recalibrated the original submillimetre data of Barvainis \etal
(1992). Our recalibration at 350$\mu m$ and 450$\mu m$ agrees with those
of Barvainis \etal (1992), whilst suggesting a flux density at 
800$\mu m$ of $66 \pm 12$\,mJy,
50\% higher than that in the original paper.
This prompted us to make a new
800$\mu m$ observation of H1413$+$117 in March 1996 which confirmed our 
recalibration with a detection of $66 \pm 9$\,mJy (R.Ivison \& J.Stevens, priv
comm.), giving a weighted mean of $66 \pm 7$\,mJy for the two
independent data sets which differs from the published $800\mu m$ 
detection of Barvainis {\it et al.} (1992) at the 2$\sigma$ level and is 
much more consistent
with that expected from a grey-body fit to the other data-points (figure 4).
The SED of the {\em Cloverleaf} is now one of the most well-defined for
a high-z object in the FIR-millimetre wavelength regime and illustrates 
the difficulty in constraining the dust temperature within the range
$30\,K < T_{d} < 70\,K$ (figure.4, see also \S 3.3.2).

\begin{figure}
\begin{picture}(220,215)
\put(0,0){\includegraphics{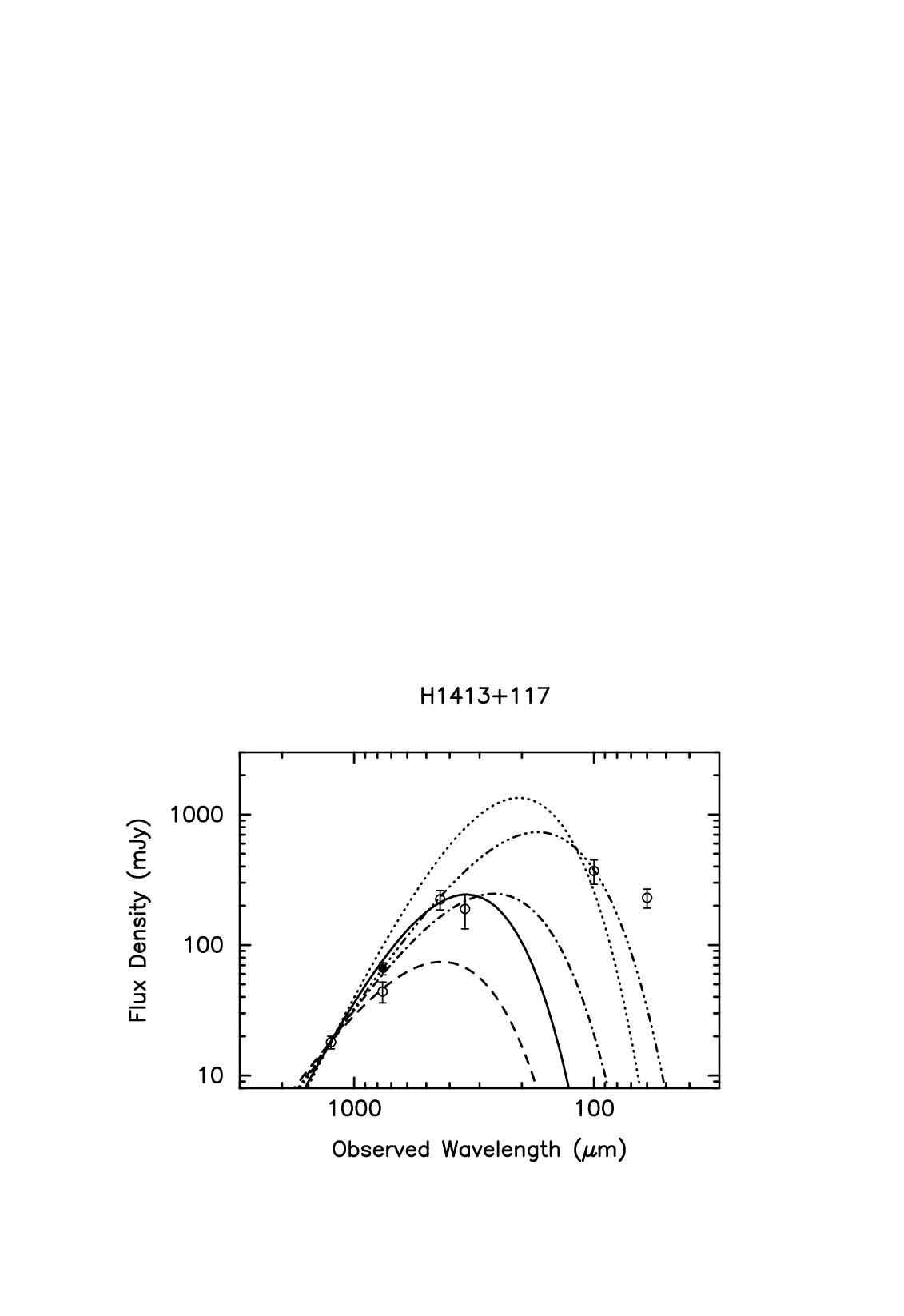}}
\end{picture}
\caption{
The mid-IR to millimetre spectral energy distribution of the {\em
Cloverleaf} quasar (H1413$+$117). The open circles show the data
presented by Barvainis {\it et al.} (1995). The solid circle shows our 
new observation at 800$\mu m$ that is consistent with various
optically-thin isothermal greybody spectra (30\,K, $\beta =2$ - solid
curve; 50\,K, $\beta =1$ - dot-dash; 76\,K, $\beta =1$ -
dot-dot-dot-dash). Also shown for completeness are 
greybodies assuming 30\,K, $\beta =1$ (dashed curve) and 50\,K, 
$\beta =2$ - dotted curve.  
The alternative greybody spectra are normalised at 1300$\mu m$. 
}
\end{figure}

\subsection{Comparison of high-z and low-z AGN at submillimetre
wavelengths}

In Table\,2 we list all  galaxies and radio-quiet quasars 
with redshifts  $z > 2$ that have  published detections 
at one or more wavelengths between $350\mu m - 1300\mu m$, regardless
of whether they have been confirmed. For the purposes of this paper
a '{\em detection}' 
is defined as any published photometry with a signal-to-noise ratio
$\ge 3$. The reliability of any detection will be discussed when appropriate.

In figure\,5 we compare the 800$\mu m$ flux densities and limits for the
high-redshift galaxies listed in Table 2 with the 800$\mu m$ flux
densities observed for a variety of starburst galaxies
and AGN at low-redshift (Hughes, Davies \& Ward 1997). 
The redshift dependence of the 800$\mu m$ flux density
of the starburst galaxy M82, previously shown in fig.\,2, is reproduced
here, together with the dependence for multiples of $\times 10$, 
$\times 100$ and $\times 1000$ the rest-frame FIR luminosity of M82 ($3
\times 10^{10} L_{\odot}$). 
This plot illustrates a number of basic but important points in a rather
transparent manner. First, with
the exception of IRAS10214$+$4624 for which, in this plot, 
we indicate the inferred 800$\mu m$ flux density after  
correcting for lensing (Eisenhardt {\it et al.} 1996),
there is an obvious absence of sources, at any redshift, 
with $S_{800\mu m} < 7$ mJy. This simply reflects the current 
sensitivity limits of sub-millimetre bolometers
such as UKT14; it is clear from the figure that our 800$\mu m$
observations of the high-z sources are amongst the faintest 800$\mu m$
observations made to date. Second, the lines on this plot indicate 
that for sources at $z > 1$, such
flux density limits correspond to FIR luminosities which are
several hundred times greater than low-z AGN. Third, this plot shows
that with an improvement in sensitivity of a factor of ten (as is
promised by new submillimetre bolometer
array detectors such as SCUBA on the JCMT) it should be possible to detect 
objects comparable to the most luminous starburst galaxies and ULIRGS 
seen in the local universe 
({\it e.g.} Arp220 and Mrk231), with luminosities L$_{FIR} \geq 
10^{12}$\,L$_{\odot}$ out to $z \simeq 10$.
We also note that little effort has been invested
in making submillimetre observations of galaxies in the intermediate  
redshift interval $z = 0.1 \rightarrow 1$.

\begin{table*}
\caption{Summary of all published  
continuum data measuring thermal emission from dust in galaxies at $z>2$
with at least one detection at wavelengths between 350$\mu m$ --
1.25\,mm. Flux density limits are quoted at the 3$\sigma$ level.} 

\begin{tabular}{lllccccc} \hline
Source name & $\,\,\,z\,\,\,$ & type &\,\,\, $S_{1.25mm}$
&\,\,\, $S_{800\mu m}$ \,\,\, & \,\,\,$S_{450\mu m}$ \,\,\, 
& \,\,\,$S_{350\mu m}$ \,\,\, & ref. \\
               &       &        & (mJy)          &   (mJy)          & (mJy)         & (mJy)         &  \\
\hline
IRAS10214+4724 & 2.286 & IRAS   &  24 $\pm$ 5    &   50 $\pm$ 5     & 273 $\pm$ 45  &  & 1 \\
53W002         & 2.390 & RG     &                &   6.9 $\pm$ 2.3  &               &  & 2 \\
H1413+117      & 2.546 & BALQSO & 18.2 $\pm$ 2.0 &   66 $\pm$ 7     & 224 $\pm$ 38  &  189 $\pm$ 56 & 2,3 \\
Q1017+1055     & 3.15  & RQQ    &  3.7 $\pm$ 1.2 &                  &  
&  & 4  \\ 
B2 0902$+$34   & 3.391 & RG     &  3.1 $\pm$ 0.6 &      $< 14$      & $< 99$              &  & 2,5 \\
4C 41.17       & 3.800 & RG     &  2.5 $\pm$ 0.4 &  17.4 $\pm$ 3.1  & $<56$        &  & 5,6 \\
PC2047+0123    & 3.800 & RQQ    &  1.9 $\pm$ 0.5 &                  &               &  & 7 \\
BR1117$-$1329    & 3.96  & RQQ  & 4.09 $\pm$ 0.81  &                  &    
&  & 4 \\
BR1144$-$0723    & 4.14   & RQQ   & 5.85 $\pm$ 1.03 &                 &
&  & 4 \\   
8C1435+643     & 4.252  & RG    &  2.6 $\pm$ 0.4 &    $< 13$        &               &  & 2,7 \\
BRI1335$-$0417 & 4.40  & RQQ    &  10.26 $\pm$ 1.04 &               & 
&  & 4 \\
BRI0952$-$0115   & 4.43  & RQQ    & 2.78 $\pm$ 0.63 &                 &
&  & 4 \\      
BR1033$-$0327    & 4.51  & RQQ  &  3.45 $\pm$ 0.65 &   12 $\pm$ 4     &               &  & 4,8,9 \\

BR1202$-$0725    & 4.69  & RQQ    & 10.5 $\pm$ 1.5 &   50 $\pm$ 7     & 92 $\pm$ 38   &  & 4,8,9 \\
\hline
\end{tabular}
\null \\
1. Rowan-Robinson {\it et al.} 1993; 2. this paper; 
3. Barvainis {\it et al.} 1995; 4. Omont {\it et al.} 1996;
5. Dunlop {\it et al.} 1994; 6. Chini \& Kr\"{u}gel 1994; 7. Ivison 1995; 
8. Isaak {\it et al.} 1994; 9. McMahon {\it et al.} 1994

\end{table*}

\begin{figure*}
\begin{picture}(200,305)
\put(0,0){\includegraphics{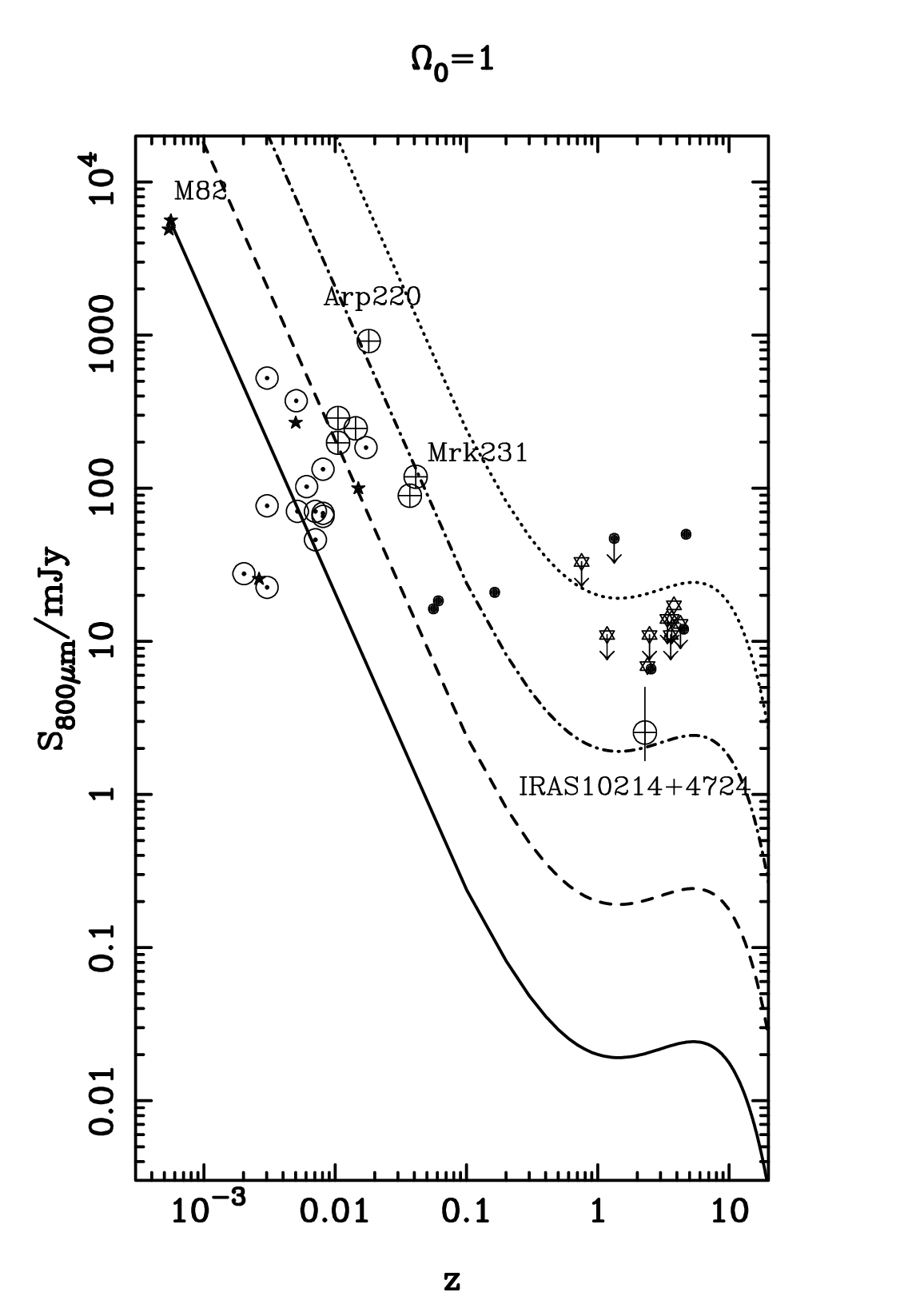}}
\put(0,0){\includegraphics{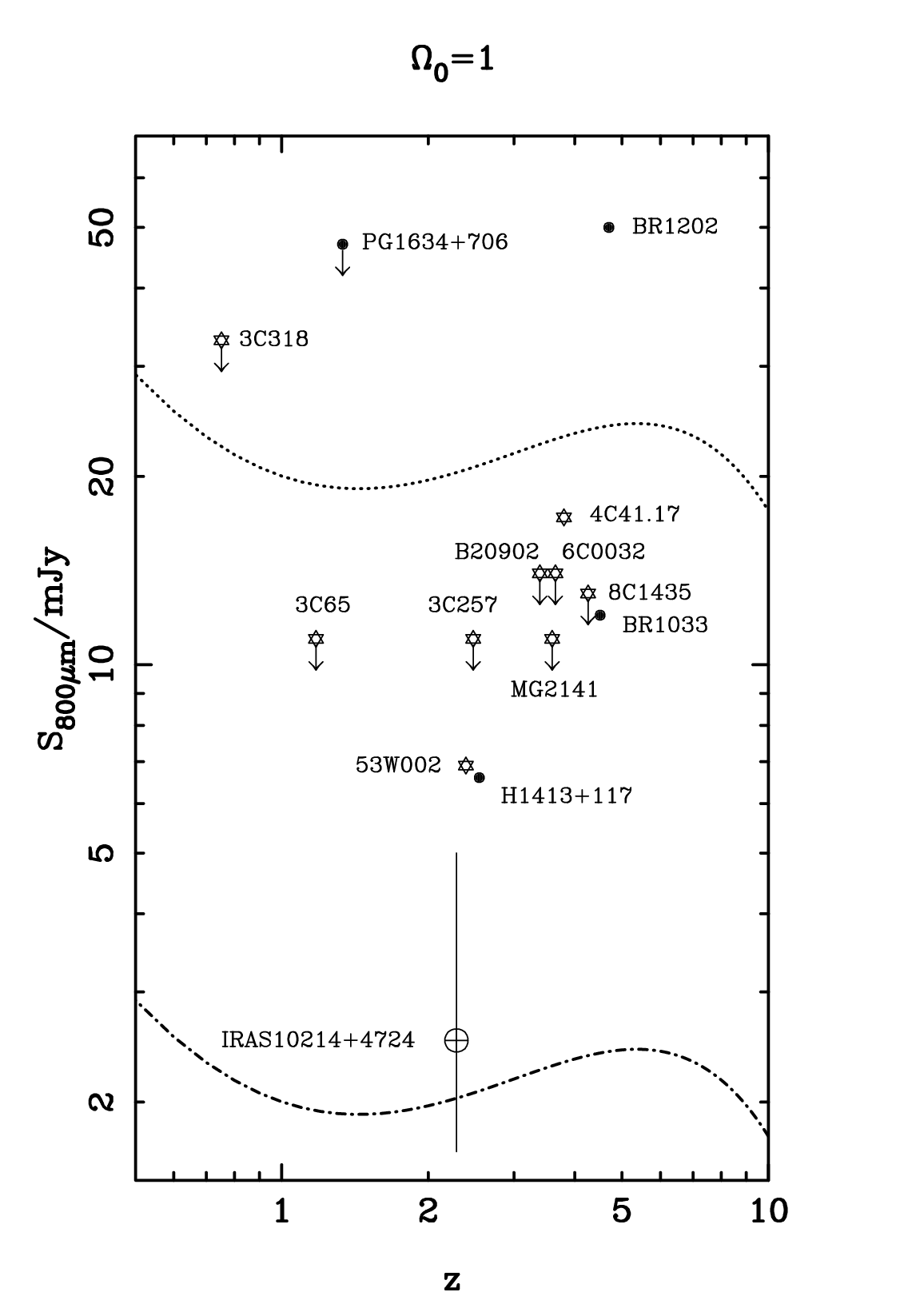}}
\end{picture}
\caption{A comparison of the measured $800\mu m$ flux densities of 
various high-z and low-z
active galaxies; starbursts (solid-stars), Seyferts (dotted-circles), ULIRGS 
(crossed-circles), radio galaxies (open-stars), radio-quiet quasars 
(solid-circles), taken from Hughes, 
Ward \& Davies (1997), Hughes {\it  et al.} (1993)
and referenced data included in Table\,2 (this paper). The flux densities of 
IRAS10214+4724 and the Cloverleaf quasar (H1413$+$117) 
have been corrected for their amplification due to lensing, see \S 4. 
The error-bar associated with IRAS10214$+$4724 represents the uncertainty
in the amplification factor at FIR wavelengths. An expanded version of
this figure between redshifts $0.5 \rightarrow 10$ is shown in the
right-hand panel.
The solid curve represents
the predicted dependence of the 800$\mu m$ flux density with redshift 
for the nearby starburst galaxy M82. Also shown are curves for a
galaxy with $\times 10$ (dashed), $\times 100$ 
(dashed-dotted) and $\times 1000$ (dotted) the rest-frame FIR 
luminosity of M82 (L$_{FIR} = 3 \times 10^{10} L_{\odot}$).}
\end{figure*}

The important physical property which can be derived directly from these
sub-millimetre detections/upper-limits is the mass of dust in a 
given galaxy. From this one can then go on to assess the evolutionary status 
of the object in question, either through estimation of the `current'
star-formation rate, or by estimating the gas mass of the object (by
adopting a given gas:dust ratio). However, in several recent papers this
crucial process has been described with misleading simplicity. In fact,
even the first stage in this
process (calculating the dust mass) has to be undertaken with great care
because there exist at least  five potential sources of significant
uncertainty. Thus, before attempting to derive the physical properties of
the objects listed in Table 2, to assess their evolutionary status
(which we do in section 4), in the next section we present a detailed
description of these 5 sources of uncertainty, assess their relative
importance and outline the prospects for improvements in each area. 

\section{Uncertainties in determining the dust masses of high-redshift objects}

In this section we describe, roughly in order of increasing subtlety, 
the problems which afflict the accurate determination of dust masses from
sub-millimetre observations of high-redshift galaxies. 
First 
we consider how one can establish that the detected emission is indeed
from dust, rather than being, for example, synchrotron emission (an
important issue for our radio galaxy targets).
Next we address
the issue of how confident one can be that it is the high-redshift target
that has been detected, and not merely foreground galactic cirrus. 
Then, assuming that it can be established that dust emission from the
high-redshift target 
has indeed been detected we consider the effect of
uncertainties both in the intrinsic properties of the dust and in its
temperature, and discuss what can be done to minimize these uncertainties. Lastly, we
briefly highlight the often overlooked implications of the current
uncertainty in the values of cosmological parameters.

\subsection{Potential Problem 1: Have we detected emission from dust?}

To prove that any sub-millimetre emission detected from high-redshift
galaxies is due to dust one must ideally demonstrate that the sub-millimetre
SED is rising too steeply to be due to self-absorbed synchrotron radiation 
({\it e.g.} Chini \etal\ 1989a, 
Hughes {\it et al.} 1993). In practice this is extremely difficult to
achieve (both due to achievable signal-to-noise,
and the fact that at high-redshift one is not always observing the
Rayleigh-Jeans tail of the expected thermal emission), but at the very
least it is necessary to demonstrate that the observed sub-millimetre
spectrum is rising towards shorter wavelengths. 

A direct determination of a positive sub-millimetre index $\alpha$ (where
$f_{\nu} \propto \nu^{\alpha}$) at high redshift is 
limited to submillimetre/millimetre 
continuum observations of only 4 objects;
H1413$+$117 (Barvainis \etal\ 1992, 1995), IRASF10214+4724 
(Rowan-Robinson {\it et al.} 1993),
4C41.17 (Dunlop {\it et al.} 1994, Chini \& Kr\"{u}gel 1994) 
and BR1202$-$0725 (Isaak {\it et al.} 1994, 
McMahon {\it et al.} 1994). 
However, in certain circumstances, even on the basis of a single
sub-millimetre or millimetre wavelength, one can be reasonably confident
that the detected emission lies above any reasonable extrapolation of the
radio emission. This point is well illustrated in Figure 6, in which we
present six examples of SEDs from the sources listed in Table 2. The SEDs 
of 4C41.17, BR1202$-$0725 and H1413$+$117 represent 
good examples of sources whose observed sub-millimetre SED is
well determined and for which there can be very little doubt that
thermal emission from dust has been detected, whilst the remaining SEDs  
illustrate the importance but also the limitations of a detection at a
single sub-millimetre or millimetre wavelength. For example, our 
800$\mu m$ detection
of 53W002, if real, appears to lie clearly above the extrapolation of its
radio spectrum, whereas the millimetre detections of B2 0902$+$34 (Chini
\& Kr\"{u}gel 1994) could simply be a detection of the high-frequency
tail of the radio emission from the unresolved core 
(Downes \etal\ 1996) which has a flatter spectrum and is 
more luminous (with respect to the submillimetre emission) than, 
for example, the radio-core in 
the ultra-steep spectrum radio galaxy  4C41.17 (Carilli \etal\ 1994).

\begin{figure*}
\begin{picture}(551,600)
\put(0,0){\includegraphics{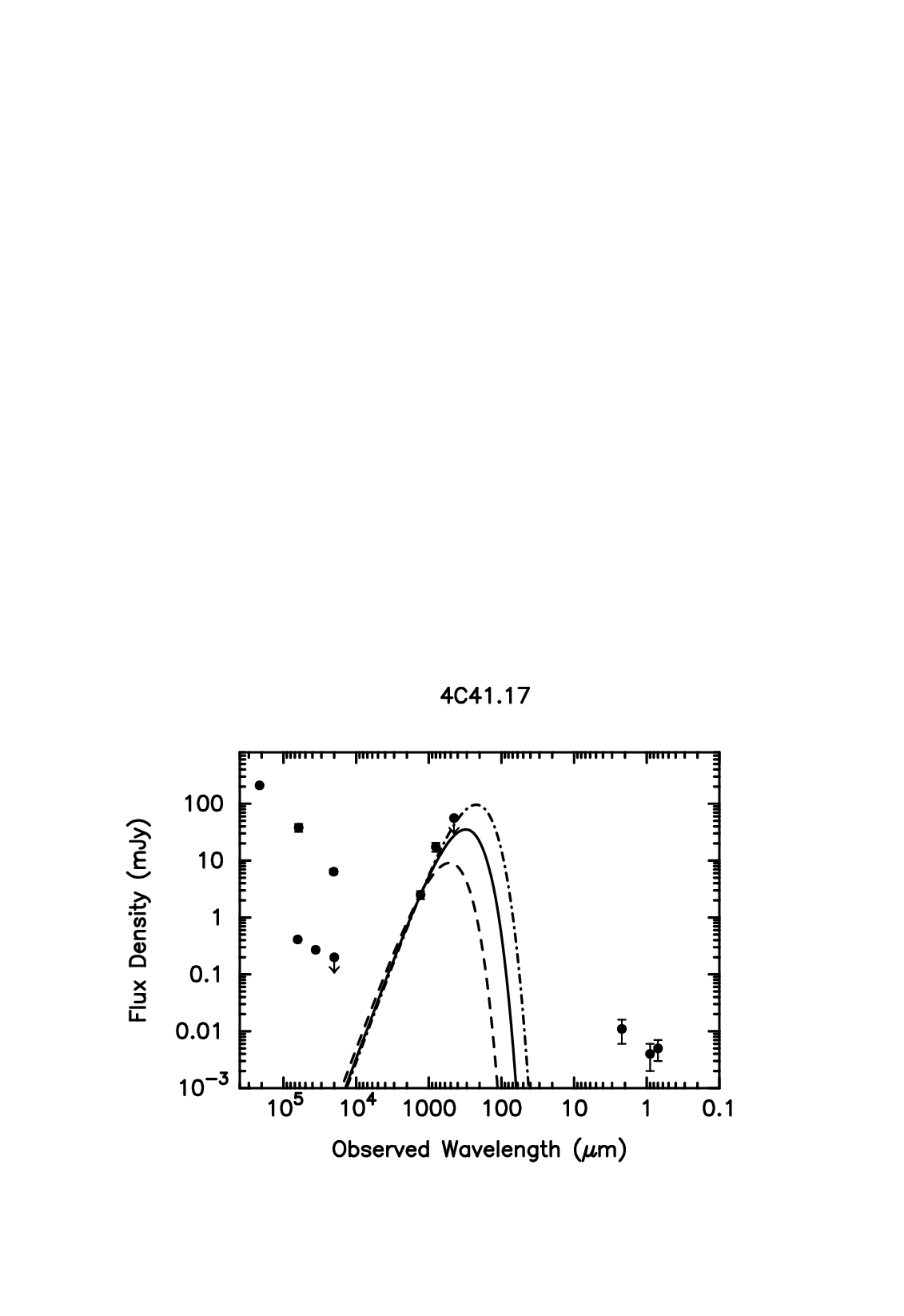}}
\put(0,0){\includegraphics{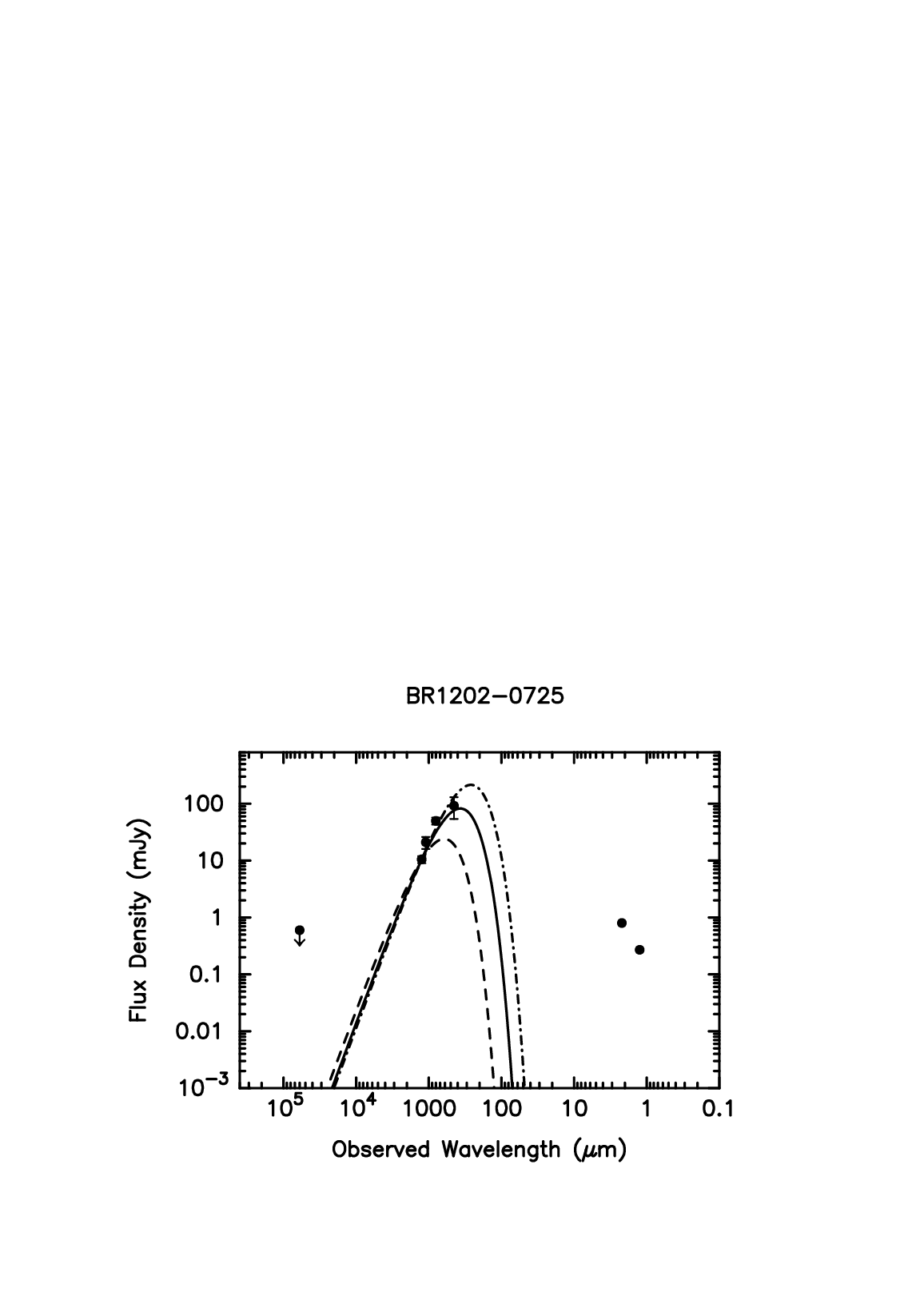}}
\put(0,0){\includegraphics{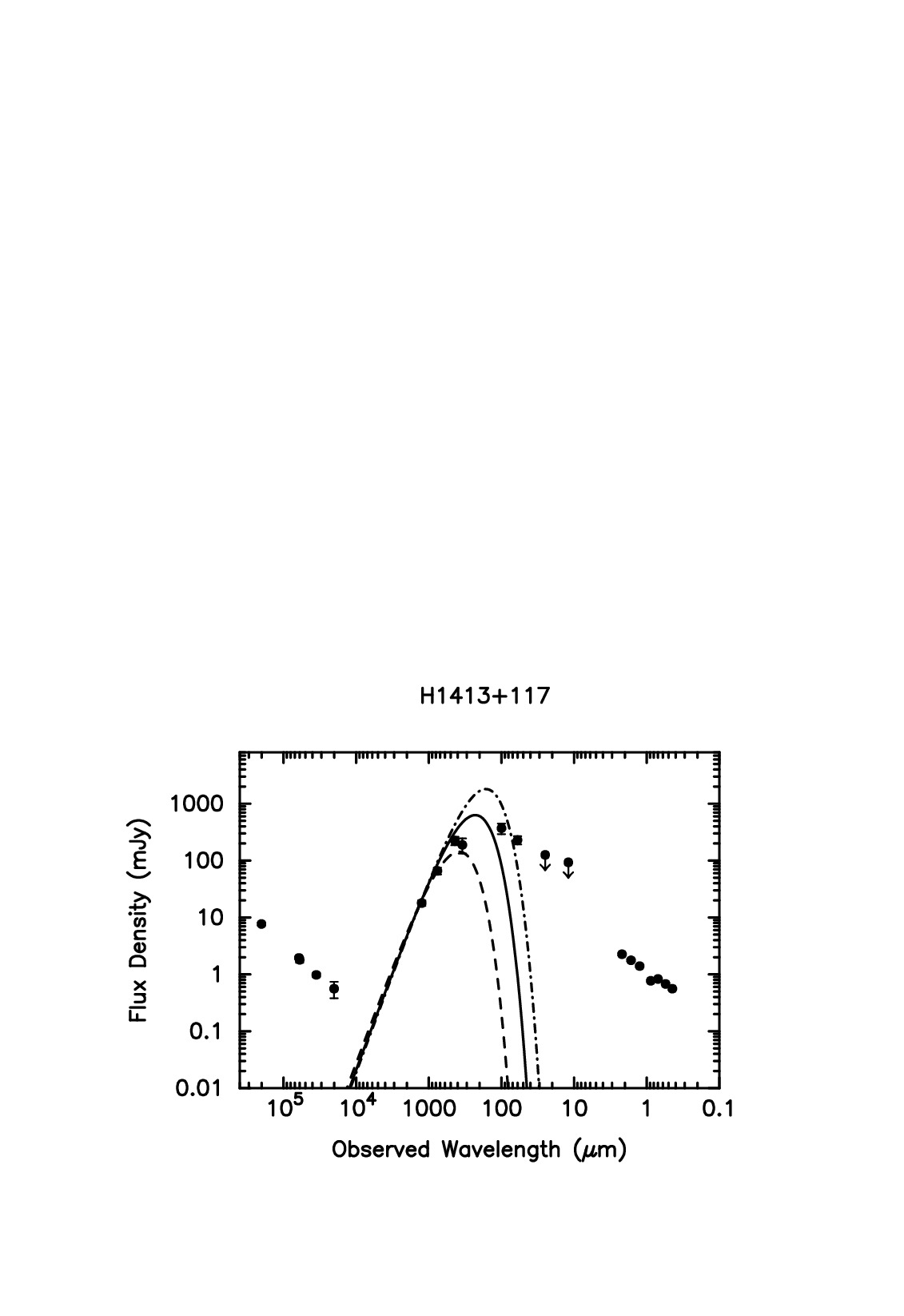}}
\put(0,0){\includegraphics{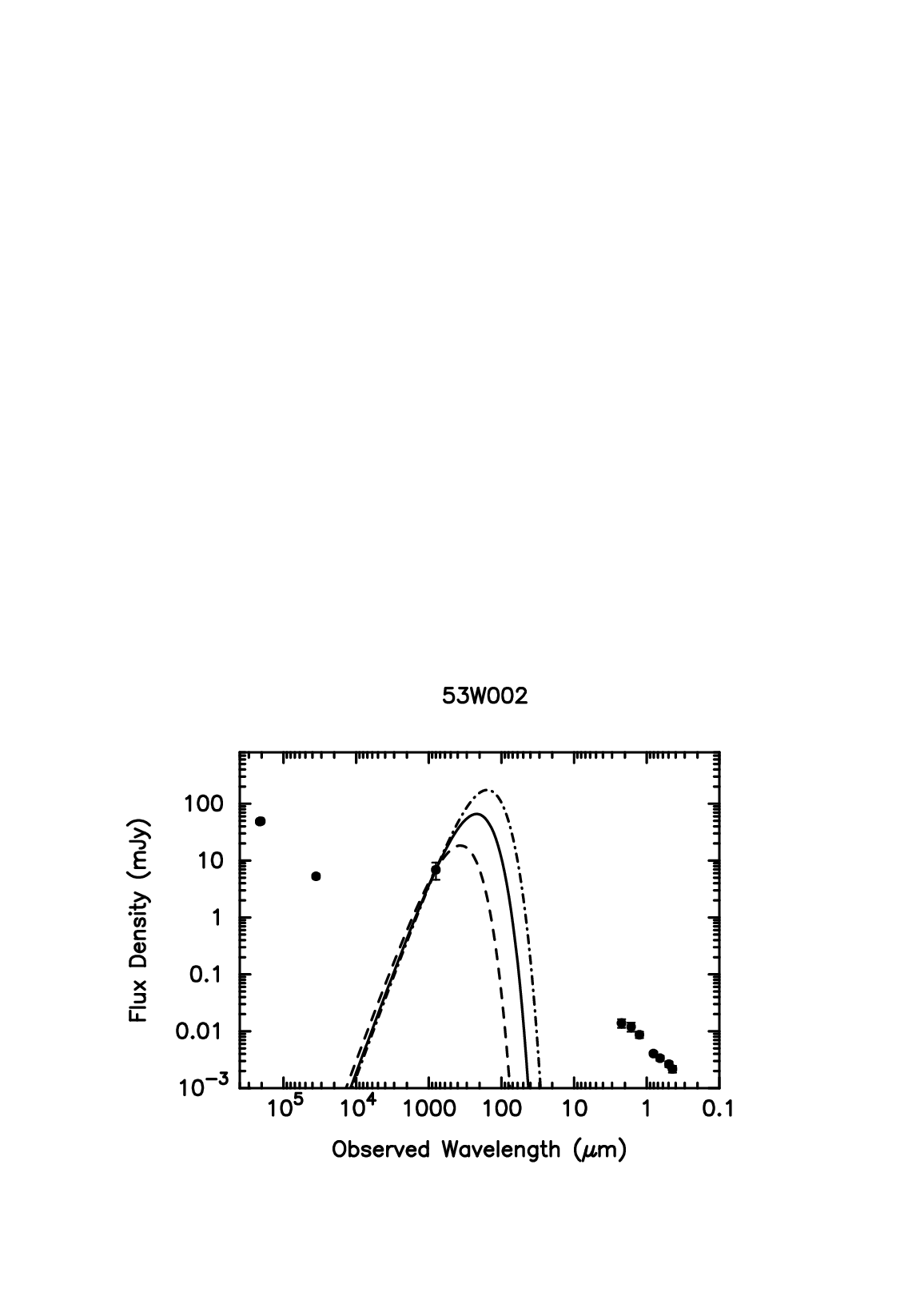}}
\put(0,0){\includegraphics{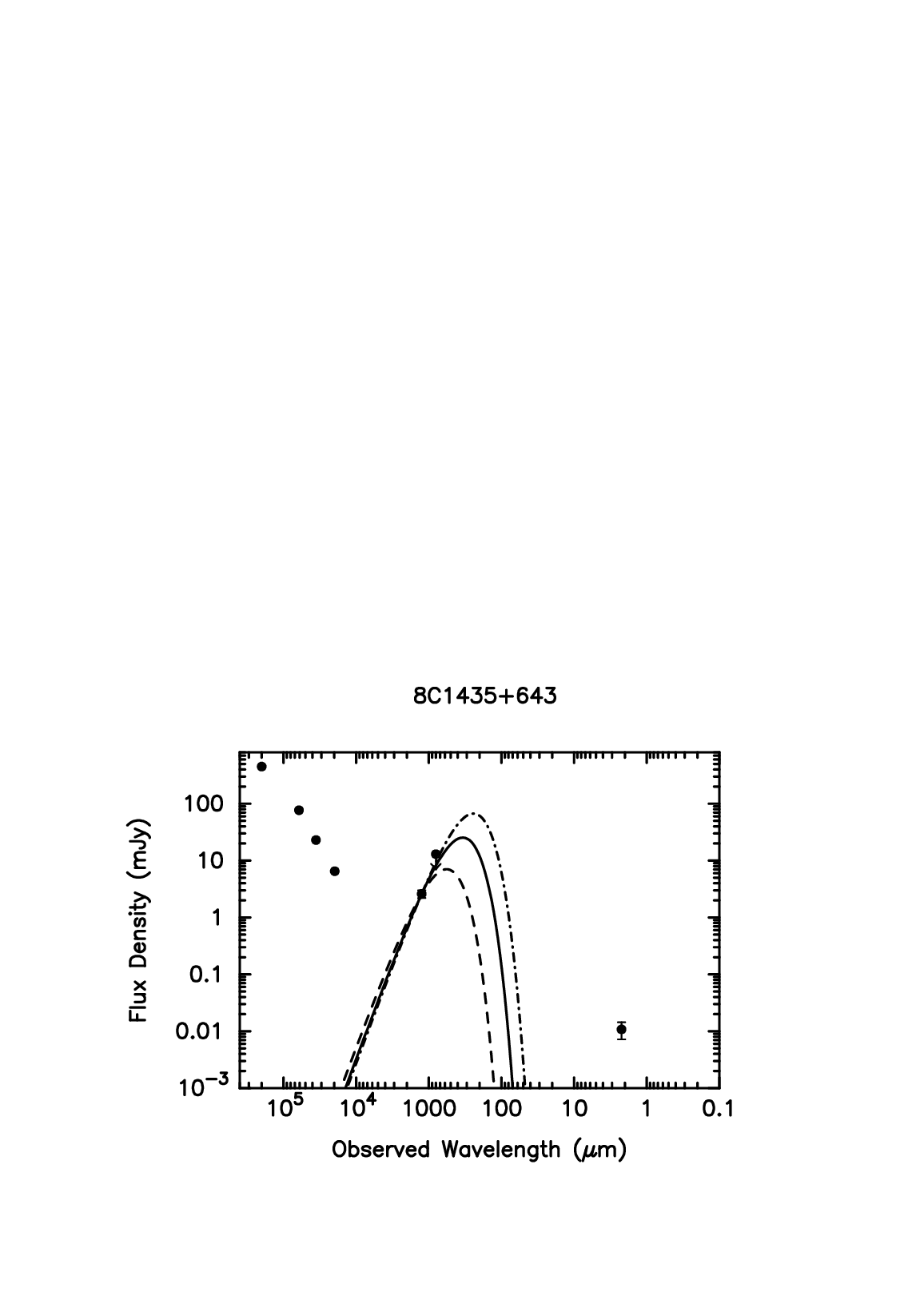}}
\put(0,0){\includegraphics{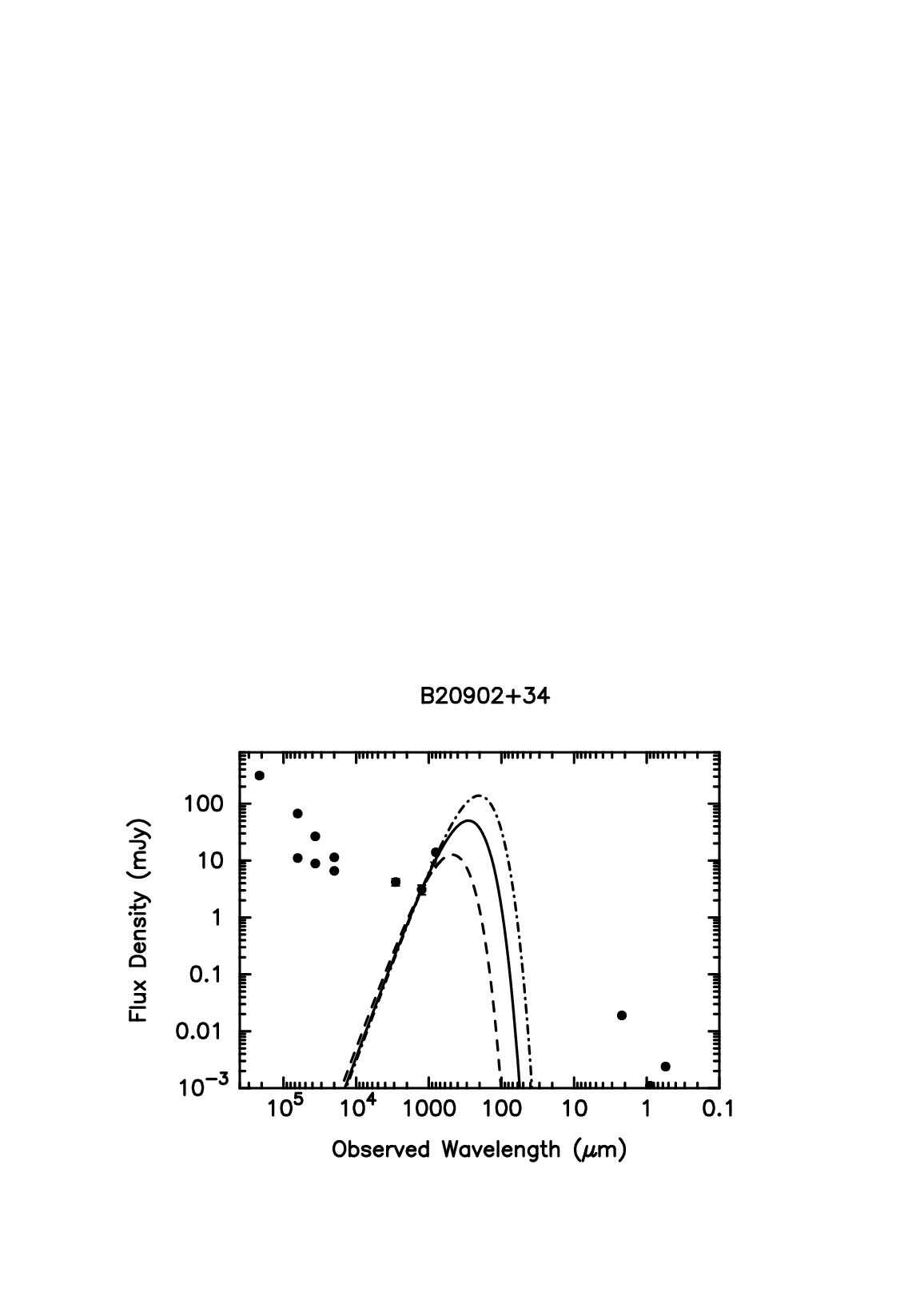}}
\end{picture}
\caption{Optical-radio SEDs of high-z radio galaxies and radio-quiet 
quasars showing that, except in the case of B20902+34 (see \S 3.1), 
the rest-frame submillimetre and FIR emission is in excess of an 
extrapolation of the total radio continuum. The SEDs of 4C41.17 and B20902+34 
show both the core emission and the steeper large-scale lobe emission. 
In each case isothermal
greybody spectra, assuming an emissivity index $\beta = 1.5$, are shown
for 30\,K (dashed), 50\,K (solid) and 70\,K (dotted-dashed). The
sub-millimetre and millimetre data are given in Table 2.
The radio, infrared and optical 
continuum fluxes for these objects have been taken
from the following papers; 4C41.17 - Carilli \etal\ (1994), Chambers \etal\
(1990: BR1202$-$0725 - Isaak \etal\ (1994):
H1413$+$117 - Barvainis \& Antonucci (1996), Barvainis \etal\ (1995): 
53W002 - Windhorst \etal\ (1991): 8C1435+643 - Lacy \etal\ (1994):
B20902+34 - Downes \etal\ (1996), Yun \& Scoville (1996), Carilli \etal\
(1994), Lilly (1988).}
\end{figure*}

Figure 6 demonstrates that with appropriate target selection ({\it e.g.}
ultra-steep spectrum radio sources) and with sensitive 
sub-millimetre and millimetre photometry one can prove that thermal
emission from dust has indeed been detected, but that this must be 
determined with care on a source-by-source basis. However, even if it can be
convincingly demonstrated that the detected emission is due to dust, one
must then address the issue of how confident one can be that the dust
lies at the redshift of the source. We therefore now address this often
over-looked but important issue.

\subsection{Potential Problem 2: What dust has been detected?: possible contamination by Galactic Cirrus}

Even if, as in the case of 4C41.17 and BR1202$-$0725, 
it can be shown that a rising sub-millimetre SED has been detected,
it must be remembered that extragalactic observations at
sub-millimetre wavelengths are made in the presence of 
foreground thermal emission from dust grains in  the 
galactic ISM  heated by the interstellar radiation field (Low \etal\ 1984),
hereafter {\em cirrus}. 
Indeed, given its fractal nature ({\it e.g.} Bazell \& D\'{e}sert 1988), 
it is possible 
that cirrus has significant structure on scales less than the typical 
$30'' - 60''$ chop-throws employed in submillimetre observations 
({\it e.g.} Meyer 1990, Diamond \etal\ 1989) 
and hence, since this foreground emission
may not be completely removed with the usual technique of position 
beam-switching, it may represent a significant source of additional noise. 
In the limit this
`{\em confusion}' noise may place a fundamental constraint on the useful
depth of searches for high-z galaxies, and it is
therefore important to quantify the contribution of
cirrus emission both to the current faint submillimetre observations 
described in this paper, and to future submillimetre high-redshift surveys.

To achieve this we have reduced the IRAS raw detector data (CRDD) at 12,
25, 60 and 100$\mu m$ at the positions of all the high-redshift objects
listed in Tables 1 \& 2, 
using the procedure described by Hughes, Appleton \&
Schombert (1991). Not surprisingly, none of the high-redshift objects 
were actually detected.  The 100$\mu m$ 
surface brightness ($I_{100\mu m}$) of the foreground cirrus was
measured towards all of the high-redshift galaxies after the
subtraction of any zodiacal emission which can still be significant in
the FIR at low ecliptic latitudes. 

The average noise level at 100$\mu m$ due to cirrus, 
on scales of a few arcminutes, was $\sim 0.3$MJy/sr  which, 
when extrapolated to 800$\mu m$, assuming an isothermal dust spectrum and
emissivity index $\beta =2$ (Rowan-Robinson 1992), gives a 
brightness of $<0.3$MJy/sr, equivalent to $<1$mJy/beam, 
for cirrus temperatures $> 12$K. In other words it is
extremely unlikely that the 800$\mu m$ detections, $> 7$\,mJy, 
in Table\,2 are due to cirrus
unless the dust grains are significantly  colder than the generally
accepted range of cirrus temperatures, $15K < T < 40K$, 
(Terebey \& Fich 1986, Rowan-Robinson 1986, Van Steenberg \& 
Schull 1988) regardless of grain-size or composition. 
However, as a caveat we mention firstly that if $\beta \neq 2$, the
submillimetre emission can increase by a factor $\sim 10$ as $\beta$ varies
between $2 \rightarrow 1$ and, secondly, that the cirrus temperatures are
generally derived from a 60$\mu m$/100$\mu m$ intensity ratio which is 
highly sensitive to the contribution from a small fraction of hot grains
in the cloud, and hence the average grain temperature that dominates the
mass of the cloud is often over-estimated. At submillimetre wavelengths 
we are significantly more sensitive to emission
from a colder component than in the FIR.  

A noise level of $\sim 1$mJy at $\sim 800\mu m$ 
is similar to the theoretical predictions of the
cirrus confusion on scales equivalent to  the
chop-throw at submm wavelengths (Helou \& Beichman 1991, Gautier, 
Boulanger, \& Puget 1992) assuming that the spatial power spectrum of cirrus,
as determined at 100$\mu m$ by IRAS (between scales of 8 degs. to 2
arcmins.), continues unbroken to smaller spatial scales ($\sim 30''$), and
that the empirical relationship between the amplitude of the variations
and the cirrus surface brightness in the FIR is valid at longer  sub-millimetre 
wavelengths. 
    
While these calculations provide reassurance that detections at the level
of those described in the present paper are unlikely to be seriously 
contaminated by cirrus emission, they also demonstrate that 
cirrus contamination is likely to 
become a more serious issue with the increased instrumental
sensitivity offered from the new generation of submillimetre receivers 
({\it e.g.} SCUBA). With such instruments it will be possible to 
reach noise levels of $\sim 0.3$mJy/beam
or 0.08\,MJy/sr in just 3 hours integration and, as a result, future 
observations of high-redshift galaxies will be sensitive to faint, hence
possibly unresolved, clumps of cirrus with
temperatures $>15$K. This result is generalised in figure 7 which 
shows the surface brightness at 800$\mu m$ of galactic cirrus
as a function of temperature. The three extrapolated models 
cover the brightness range of cirrus at 100$\mu m$ 
($0.1\,MJy/sr < I_{100\mu m} < 10.0\,MJy/sr$) 
towards the positions of the  high-redshift galaxies given in Tables 1
\& 2 and are typical of the cirrus background away from the galactic
plane ($b > 10^{o}$) in all but the darkest regions ({\it e.g.} Lockman
\etal\ 1996).

\begin{figure}
\begin{picture}(350,250)
\put(0,0){\includegraphics{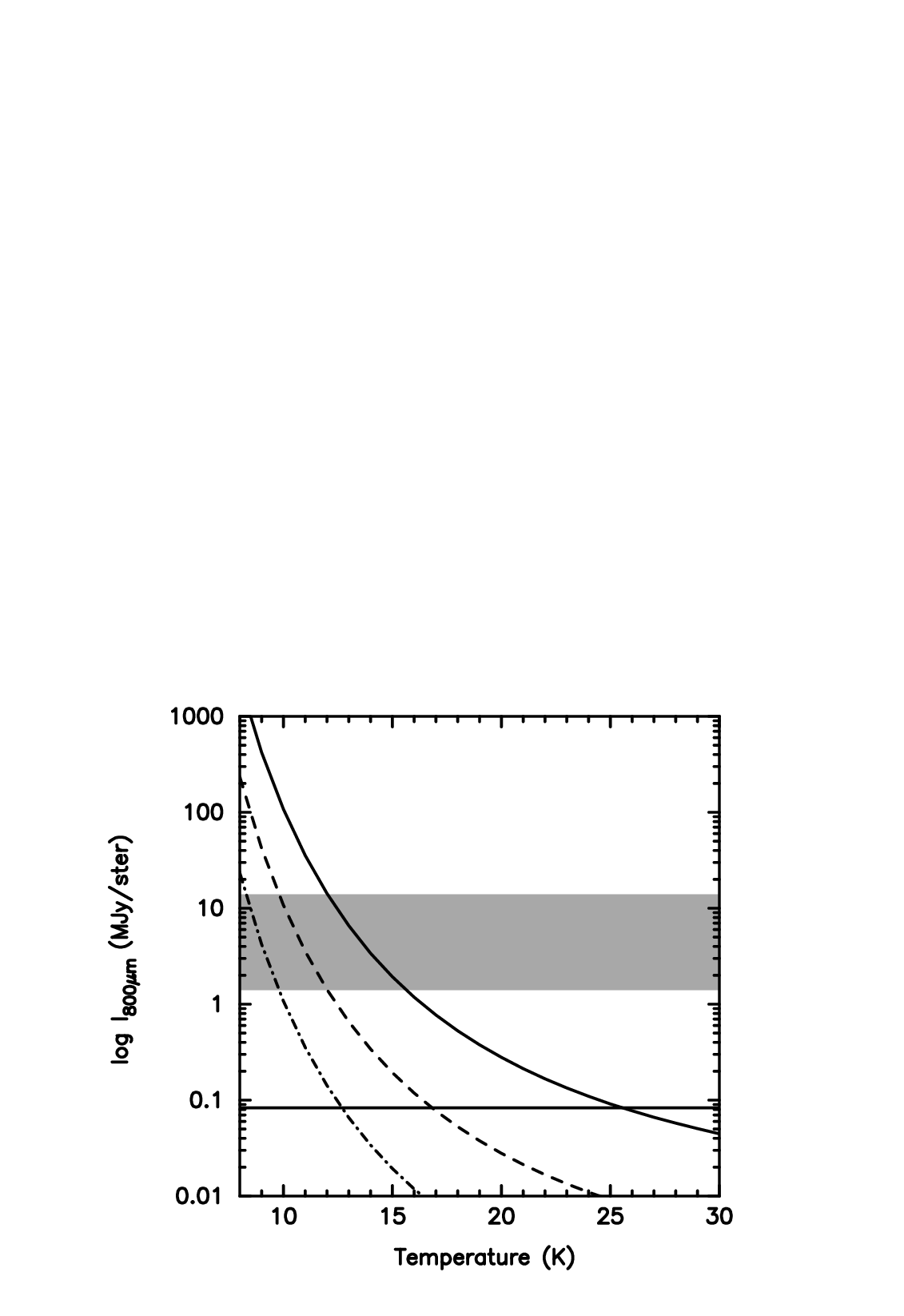}}
\end{picture}
\caption{Surface brightness at 800$\mu m$ of galactic cirrus as a function
of dust temperature. The curves are extrapolations of 
optically-thin, isothermal models
with a  grain emissivity index $\beta = 2$, assuming a 100$\mu m$ intensity of 
10\,MJy/ster (solid line), 1\,MJy/ster (dashed-line), 0.1\,MJy/ster
(dot-dashed line). 
The shaded region represents the range of the surface brightness of the 
800$\mu m$ detections in Table\,2 and illustrates that for
$T \gg 15$K the contribution from cirrus is negligible.
The horizontal line represents the intensity level at $\sim 800\mu m$ 
that can
be reached in 3 hours integration with the future generation of submillimetre
receivers, and indicates that deep cosmological studies may suffer from
source confusion due to galactic cirrus emission. } 
\end{figure}

\subsection{Potential Problem 3: Uncertainties in the properties and
temperature of the dust grains}

If it is assumed the submillimetre continuum ($\lambda_{rest} > 200\mu m$) 
is due to optically-thin emission from heated dust grains with no 
additional contribution from bremsstrahlung or synchrotron radiation, a 
measure of the dust mass $M_{d}$ can be determined directly from the 
relationship,

\be M_{d} = \frac{1}{1+z} \frac{S_{obs} D_{L}^{2}}{k_{d}^{rest}
B(\nu^{rest},T)} \ee

\noindent 
where $z$ is the redshift of the source,
$S_{obs}$ is the observed flux 
density, 
$k_{d}^{rest}$ is the rest-frequency mass absorption coefficient,
$B(\nu^{rest},T)$ is the rest-frequency value of the Planck function
from dust grains radiating at a temperature $T$, and $D_L$ is the luminosity
distance (see section\,3.4).

Thus, for a given cosmology and excluding the measurement uncertainty of the
continuum flux density, the robustness of dust mass 
determinations from 
submillimetre photometry depends on the uncertainty in $k_d$ and $T$.
The existing constraints on these two parameters are now briefly 
discussed in turn.

\subsubsection{Uncertainty in $k_d(\lambda)$}

Draine (1990) has pointed out that the acknowledged advantage 
of using optically-thin  submillimetre emission to determine the dust mass 
in galaxies is 
offset somewhat by increased uncertainty in the properties 
of interstellar dust, and hence the value of $k_d$ as $\lambda$ 
is increased from FIR to sub-millimetre wavelengths.
A reasonable estimate of the maximum 
fractional uncertainty in $k_d$ at 800$\mu m$
is $\simeq 7$, with the values of $k_d(800\mu m)$ ranging between 
0.04 m$^2$ kg$^{-1}$ (Draine \& Lee 1984) and 
0.3 m$^2$ kg$^{-1}$ (Mathis \& Whifffen 1989) with intermediates values of
0.15 m$^2$ kg$^{-1}$ (Hildebrand 1983) and 0.12$m^{2} kg^{-1}$
(Chini, Krugel \& Kreysa 1986). In order that we can extrapolate $k_{d}$
to shorter rest-wavelengths, appropriate for high-z galaxies,
we have assumed that $k_{d} \propto \lambda^{-1.5}$ and  have 
adopted an average value of $k_d(800\mu m) = 0.15 \pm 0.09$\, 
m$^2$ kg$^{-1}$.
A different choice of $k_d$ should therefore 
only be expected to result in dust mass 
estimates which differ from those which we calculate in section 4, 
by at most a factor of $\simeq 2$. 

\subsubsection{Uncertainty in $T$}

The temperature of dust grains that radiate at submillimetre wavelengths
and dominate the FIR luminosity in nearby starburst galaxies, low-metallicity
dwarf galaxies, ULIRGs,
Seyferts and radio-quiet quasars is typically $50 \pm 20$\,K 
(Chini \etal\ 1989a, Chini \etal\ 1989b, 
Hughes {\it et al.} 1993, Hughes, Davies \& Ward 1997).
There is no {\em a priori} reason to believe that the dust radiating at
FIR rest wavelengths in high-redshift radio galaxies and radio-quiet quasars
should be significantly different if the star formation processes that
exist at early epochs are similar to those in the local universe.

The fractional uncertainty in the dust mass which results
from ignorance of the dust temperature within the range
$T_{1} < T < T_{2}$ is given by

\be \frac{M_1}{M_2} =
\frac{e^{h\nu_{rest}/kT_1}-1}{e^{h\nu_{rest}/kT_2}-1} \ee

\noindent 
which increases rapidly as $\nu_{rest}$ moves above the Rayleigh-Jeans tail of
the thermal dust emission. Dust mass estimates derived 
from {\it rest-frame} sub-millimetre photometry therefore have the 
benefit of being relatively insensitive to 
uncertainties in temperature, as compared to dust masses 
derived from far-infrared data ({\it i.e.} $\lambda_{rest} < 200\mu m$).

At $z \simeq 0$, where 800$\mu m$ photometry samples the 
Rayleigh-Jeans tail of the thermal dust emission for T$_{d} \gg 20\,K$, 
the fractional
uncertainty in the estimated dust mass is
proportional to the fractional uncertainty in the dust temperature, but 
as figure 8 illustrates, the uncertainty in $M_{dust}$ increases sharply with increasing redshift
unless the observing frequency is reduced appropriately.
For example,  at $z \simeq 3$ the difference in dust mass between
assuming $T = 30\,K$ and $T = 70\,K$ is a factor of 15 and 5
 at 450$\mu m$ and 800$\mu m$ respectively.  
Consequently our ignorance of the most appropriate value of $T_{dust}$ 
in high-redshift galaxies  is 
a dominant source of uncertainty in the derived dust masses
and it is worth considering the implications 
of such an uncertainty in future cosmological
studies at submillimetre wavelengths.

\begin{figure}
\begin{picture}(220,250)
\put(0,0){\includegraphics{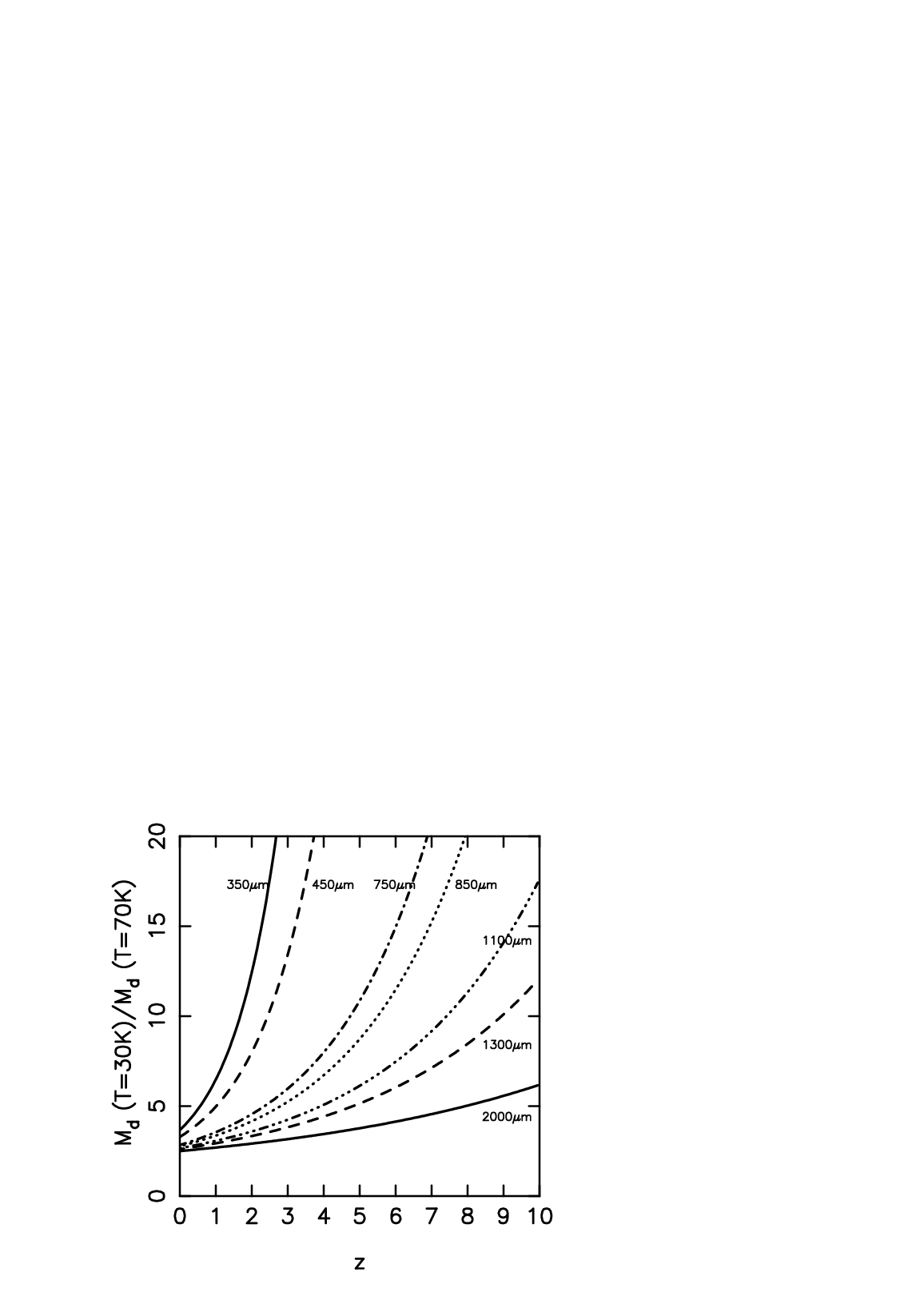}}
\end{picture}
\caption{Temperature-dependent uncertainty 
in dust masses deduced from sub-millimetre photometry in various 
wavebands plotted as a function of source redshift.
If we consider a fractional error in dust mass of greater than 
5 to be unacceptable, then photometry at $800\mu m$ is 
of little use for observations of galaxies at $z > 4$, 
while photometry at $\simeq 400\mu m$
is of little worth (on its own) for $z > 1$.}
\end{figure}

It is clear that the more one benefits from the increased flux levels offered
by intercepting the thermal spectrum nearer the rest-frame 
$100\mu m$ peak, the greater uncertainty there is
in the derived dust mass determined from a single submillimetre
continuum measurement. As we have summarised in Table 2, 
almost all photometric 
observations at millimetre and submillimetre wavelengths of high-z galaxies 
have been restricted to filters at  $\lambda \geq 800\mu m$,
with the exception of IRAS10214$+$4724, 4C41.17, H1413$+$117 and BR1202$-$0725. 
Thus the majority of data have been made in the rest-frame Rayleigh-Jeans 
regime (for $T>30$\,K) and consequently, despite the inability of the data to 
constrain the temperature, the uncertainty in the mass due to this effect 
alone is no worse than a factor of 5 for galaxies at a redshift $z < 4$.

Figure\,8 clearly demonstrates the value of a general philosophy 
when determining the dust mass from  a single photometric detection,
that is to use the longest available
submillimetre or millimetre wavelength particularly for galaxies at $z > 1$.

\subsection{Potential Problem 4: Uncertain cosmology}

The luminosity distance, $D_L$, which appears in equation\,2 is given by 

\be D_{L} = \frac{2c}{H_0 \Omega_0^2} \left\{\Omega_0 z + (\Omega_0 -
2)[(\Omega_0 z + 1)^{1/2} - 1] \right\}\ee

The (apparently now decreasing) uncertainty in the value of the Hubble
constant $H_0$ obviously affects the calculated values of dust mass to
some extent, but
comparison of the properties of low and high-redshift galaxies is
unaffected because $M_{dust}$ simply scales as $H_0^{-2}$.

More serious is our current ignorance of $\Omega_0$.
At $z \simeq 4$ if one adopts a low density universe ({\it i.e.}
$\Omega_0 \simeq 0.1$) the derived dust masses are a factor $\simeq 5$
greater than if one assumes an Einstein de-Sitter Universe, and this 
uncertainty factor rises to $\simeq 15$ at redshifts $z \simeq 10$.
The value of pointing out the size of this
uncertainty is that it puts into perspective the relatively minor
errors which are introduced by our ignorance of the precise value of such
parameters as $k_d$.

Bearing these uncertainties in mind, we now proceed to calculate our best
estimates of the dust masses of the high-redshift objects listed in Table
2, and hence to infer their evolutionary status.

\section{Derivation of physical parameters for high-redshift galaxies}

\subsection{The dust masses of high-redshift galaxies}

At first sight the preceding section may read as a rather depressing
litany of ever-increasing sources of error. However, our objective is not
to denigrate the usefulness or importance of sub-millimetre cosmology, but
rather to delineate and quantify the main sources of error, and to identify 
what strategies must be adopted to maximise the usefulness of
sub-millimetre and millimetre continuum photometry. In fact, for most of
the sources listed in Table 2, the current observational constraints
indicate that we can still derive realistic and meaningful estimates of
their dust masses, with appropriate choices of parameters. First, as
illustrated in figure 6, for most of these sources we can be  
confident that the detected sub-millimetre emission is produced by dust.
Second, while it appears that cirrus confusion may be a significant
problem for deeper surveys, our estimates of the background cirrus noise
for the sources detected here indicate that these detections are unlikely
to be significantly effected. Third, we can minimize the effect of 
disagreement of $k_d$ by choosing an average value. Fourth, 
for at least some sources ({\it e.g.} 4C41.17, BR1202$-$0725, H1413$+$117) 
there exists sufficient observational
data to set a meaningful upper limit on the dust temperature, and hence
lower-limit on the dust mass.
Fifth, while it is obviously beyond the scope of this paper to address
the issue of the true value of $\Omega_0$, by adopting a high-density
universe we can at least ensure that our dust mass estimates are
conservative.

We have therefore calculated dust masses for the sources listed in Table\,2 
using equation\,2, assuming $k_d (800\mu m)  = 0.15$m$^{2}$kg$^{-1}$, 
$\beta = 1.5$, $T_{dust} = 50$K and
$\Omega_0 = 1$. The results are presented in Table\,3  and are
in the range $1 \times 10^{8} - 2 
\times 10^{9} M_{\odot}$ allowing for amplifications  of order $\sim 11$
(Barvainis {\it et al.} 1994) 
and $\sim 10-30$ (Eisenhardt {\it et al.} 1996) for the FIR luminosity 
in the lensed sources H1413$+$117 and IRAS10214$+$4724 respectively. 

These dust mass estimates can be used to infer the evolutionary status of
the galaxies in two alternative ways. First one can estimate the
`current' star-formation rate of the galaxy. Second one can attempt to
estimate the amount of molecular gas in the galaxy which remains to be
converted into stars at the epoch of observation. We now consider each of
these calculations in turn.

\begin{table*}
\caption{Dust masses, molecular gas masses,
 FIR luminosities and starformation rates (SFRs)
of high-z galaxies.
Corrections for the amplification of the FIR emission
due to lensing have been applied to the values for IRAS10214$+$4724
and H1413$+$117 (see \S 4.1).} 

\begin{center}
\begin{tabular}{llrrrr} \hline
Source name & $\,\,\,z\,\,\,$ & log M$_{d}$/M$_{\odot}$ & 
log M$_{H_{2}}$/M$_{\odot}$ & log L$_{FIR}$/L$_{\odot}$ & SFR (M$_{\odot}$/yr) \\
\hline
3C318          & 0.752 \hspace{3mm} & $< 9.16$ \hspace{5mm} & $< 11.86$ \hspace{5mm} & $< 13.78$ \hspace{5mm} & $< 6025 $ \hspace{5mm} \\
3C65           & 1.176 \hspace{3mm} & $< 8.72$ \hspace{5mm} & $< 11.42$ \hspace{5mm}  & $< 13.35$ \hspace{5mm} & $< 2239 $ \hspace{5mm} \\
PG1634$+$706   & 1.334 \hspace{3mm} & $< 9.35$ \hspace{5mm} & $< 12.04$
\hspace{5mm} & $< 13.97$ \hspace{5mm} & $< 9332 $ \hspace{5mm} \\   
IRAS10214$+$4724 
               & 2.286 \hspace{3mm} & 8.02 \hspace{5mm}     & 10.72 \hspace{5mm}   & 12.64 \hspace{5mm} & 436 \hspace{5mm}   \\
53W002         & 2.390 \hspace{3mm} & 8.45 \hspace{5mm}     & 11.15 \hspace{5mm} & 13.08 \hspace{5mm} & 1202 \hspace{5mm}  \\
3C257          & 2.474 \hspace{3mm} & $< 8.65$ \hspace{5mm} & $< 11.35$ \hspace{5mm} & $< 13.27$ \hspace{5mm} & $< 1862 $ \hspace{5mm} \\
H1413$+$117    & 2.546 \hspace{3mm} & 8.42 \hspace{5mm}     & 11.12 \hspace{5mm}    & 13.04 \hspace{5mm}    & 1096 \hspace{5mm}  \\
Q1017+1055     & 3.15 \hspace{3mm} & 8.67 \hspace{5mm}     &   11.37
\hspace{5mm}  & 13.29 \hspace{5mm}  & 1949 \hspace{5mm} \\    
B20902$+$34   & 3.391 \hspace{3mm} & 8.61  \hspace{5mm}    & 11.31 \hspace{5mm}    & 13.24 \hspace{5mm}    & 1738 \hspace{5mm}  \\
MG2141$+$192   & 3.594 \hspace{3mm} & $< 8.58$ \hspace{5mm} & $< 11.28$ \hspace{5mm} & $< 13.20$ \hspace{5mm} & $< 1585$ \hspace{5mm} \\
6C0032$+$412   & 3.650 \hspace{3mm} & $< 8.68$ \hspace{5mm} & $< 11.38$ \hspace{5mm} & $< 13.30$ \hspace{5mm} & $< 1995$ \hspace{5mm} \\
4C41.17        & 3.800 \hspace{3mm} & 8.76 \hspace{5mm}     & 11.46 \hspace{5mm}    & 13.39 \hspace{5mm}    & 2455 \hspace{5mm}  \\
PC2047$+$0123  & 3.800 \hspace{3mm} & 8.36 \hspace{5mm}     & 11.06
\hspace{5mm}    & 12.99 \hspace{5mm}    & 977 \hspace{5mm}  \\
BR1117$-$1329  & 3.96 \hspace{3mm} & 8.62 \hspace{5mm} & 11.32 \hspace{5mm} & 13.24 \hspace{5mm} &
1737 \hspace{5mm} \\
BR1144$-$0723  & 4.14 \hspace{3mm} & 8.77 \hspace{5mm} & 11.47
\hspace{5mm}  & 13.40 \hspace{5mm} & 2511 \hspace{5mm}  \\  
8C1435$+$643   & 4.252  \hspace{3mm} & 8.46 \hspace{5mm}     & 11.16
\hspace{5mm}    & 13.08 \hspace{5mm}    & 1202 \hspace{5mm}  \\
BRI1335$-$0417 & 4.40 \hspace{3mm} & 8.97 \hspace{5mm} & 11.67
\hspace{5mm} & 13.61 \hspace{5mm} & 4074 \hspace{5mm} \\  
BRI0952$-$0115 & 4.43 \hspace{3mm}   & 8.43 \hspace{5mm} & 11.13
\hspace{5mm} & 13.05 \hspace{5mm} & 1122 \hspace{5mm}    \\
BR1033$-$0327  & 4.51  \hspace{3mm} & 8.57 \hspace{5mm}     & 11.27 \hspace{5mm}    & 13.19 \hspace{5mm}    & 1549 \hspace{5mm}  \\
BR1202$-$0725  & 4.69  \hspace{3mm} & 9.18 \hspace{5mm}     & 11.88 \hspace{5mm}    & 13.81 \hspace{5mm}    & 6456 \hspace{5mm} \\
\hline
\end{tabular}
\end{center}

\end{table*}

\subsection{Star formation rates in high-redshift galaxies}

As described in the introduction, one can estimate the star-formation
rate at the time of observation from their rest-frame FIR
luminosities which have been calculated between rest-wavelengths 
of $2\,mm \rightarrow 10\mu m$ assuming an optically-thin isothermal 50K
greybody spectrum (see Table\,3), which can be reasonably justified in
the case of 4C41.17, BR1202$-$0725 and H1413$+$117 
(figure\,6, see also \S 4.2.1), normalised to  the millimetre and 
submillimetre flux densities in Table 2. In the absence of any 
contribution from an AGN or amplification due
to lensing, the rest-frame FIR luminosities, which lie in the range
$4 \times 10^{12} \rightarrow 6 \times 10^{13}$\,L$_{\odot}$,
imply large current SFR's of greater than several $100 M_{\odot}$/yr. 
A comparison of the $1000 - 8\mu m$ luminosities of low-z ULIRGs (Sanders \etal\
1991) with those calculated from a single 50\,K isothermal model indicates
that we have underestimated the rest-frame FIR luminosities of the
high-z galaxies by a factor of $\sim 2.5$ if they contain  
contributions from hotter dust ($T> 100$K) components which radiate
strongly at mid-IR wavelengths. Hence the FIR luminosities and SFRs 
in Table\,3 should be treated as lower-limits when compared to those 
of lower-redshift galaxies. Further, albeit controversial
evidence for extreme SFRs and young galaxy ages ($< 1$\,Gyr) have
been found in the rest-frame UV-optical morphologies and SEDs
of 53W002 and 4C41.17 (Windhorst {\it et al.} 1991, Chambers {\it et al.} 1990,
Mazzei \& de Zotti 1995).

Thus, despite the uncertainties described in \S 3 
we can still  conclude that the high-z radio galaxies and 
radio-quiet quasars, which have been detected at submm-mm wavelengths, 
are extremely dusty, with dust masses 
$> 10\times$ larger than observed in their low-z ($z<0.5$) counterparts 
(Chini {\it et al.} 1989a, Knapp \& Patten 1991, Hughes {\it et al.} 1993),
and with levels of starformation and starforming efficiencies 
similar to, or exceeding those observed in low-z ULIRGS (see figure\,9). 

\begin{figure*}
\begin{picture}(345,335)
\put(0,0){\includegraphics{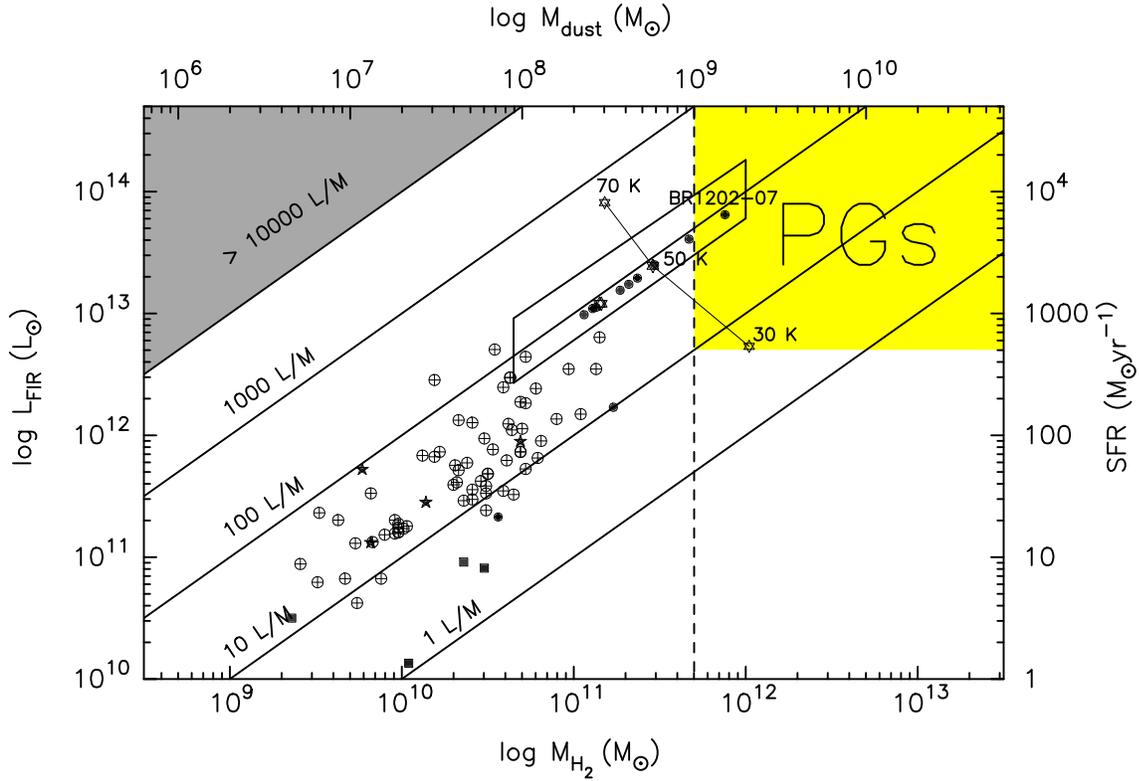}}
\end{picture}
\caption{A representation of the data given in Table\,3 for the
detected high-z radio galaxies and radio-quiet quasars, 
compared to their lower-redshift counterparts, ULIRGs and
elliptical galaxies. The symbols have the 
same meaning as in figure\,5, with additional low-z powerful radio
galaxies (Mazzarella \etal\ 1993), ULIRGs (Sanders \etal\ 1991) and 
elliptical galaxies (Lees \etal\ 1991)
detected in CO marked as solid stars, crossed-circles and solid squares 
respectively.  
The diagonal lines indicate constant L$_{FIR}$/M$_{H_{2}}$, while
the vertical dashed-line shows the gas mass
boundary, to the right of which one can expect to find the projenitors
of the most massive elliptical galaxies. The parallelogram encloses the high-z galaxies, which lie along the
lower boundary, and depicts the typical range of the  
increase in their FIR luminosities and SFRs due to a 
contribution from hotter ($T>50$K) dust (see \S 4.2). 
This figure demonstrates that it is
difficult, except possibly in the case of BR1202$-$0725 (although it may
be lensed, see \S 5), to describe any of the 
high-z galaxies detected at submm and mm wavelengths as genuinely {\em
prim\ae val}, since they lie outside the shaded region marked "PGs" 
which represents the parameter space populated by
primaeval galaxies, with our definition of M$_{H_{2}} > 5 \times
10^{11}$\,M$_{\odot}$ and a SFR\,$> 500$\,M$_{\odot}$.  
Varying the assumed dust temperature through a reasonable range 
($30K \rightarrow 70\,K$), as illustrated by the representative
locus passing through 4C41.17, struggles to change this basic conclusion,
except perhaps for 4C41.17 itself.
}
\end{figure*}

Particularly if regarded as lower limits, these inferred SFRs are 
sufficiently high to be at least consistent with the formation of a 
giant elliptical galaxy in $\simeq 1$ Gyr. However, such a
calculation is open to two important criticisms. First, for the active
objects considered here it is unclear whether the dust is heated primarily
by young stars or by a quasar nucleus, although this argument can be at
least partially countered by the fact that the very existence of such
large masses of dust indicates extensive recent star-formation. A second,
and more important criticism is that an estimate of the `current' SFR 
can never shed any light on the length of time over which 
such a high star-formation rate has been maintained. In other words it is
clearly impossible to distinguish a brief, albeit violent burst of
star-formation activity (triggered, perhaps, by an interaction) from the
sort of sustained high-level star-formation required to produce $\simeq
10^{11} \rightarrow 10^{12} M_{\odot}$ of stars. What is required is some
estimate of the past and future star-formation history of the galaxy.
Such an estimate can be provided, to first order, by using the calculated
dust mass of a galaxy to estimate its gas mass.

\subsubsection{Are simple isothermal models adequate?}
We have assumed that the rest-frame FIR 
luminosities in table\,3 are dominated by radiation from grains with an average 
temperature of 50K, and that 
the SFRs can be derived from a single isothermal component.
This is of course a simplification, since {\em real} galaxies clearly 
have a distribution of grain temperatures that contribute to the 
rest-frame emission between 200$\mu m$ - 50$\mu m$. 
However the thermal emission from grains which radiate at a
temperature $T$, with an
efficiency Q$_{abs} \propto \nu^{\beta}$, has a peak in the spectrum of
$S_{\nu}$ at a wavelength $\lambda (\mu m) \sim \frac{5100}{T(K)} \left(
\frac{3}{3+\beta} \right)$, and consequently there remains only
a restricted range of temperatures (20--70\,K)
that contribute significantly to the
the rest-frame FIR. 

In figure 10 we quantify the discrepancy between our calculations of the
FIR luminosities in table\,3 and, where overlap exists, the bolometric
starburst luminosities calculated from the more sophisticated 
radiative transfer models of Rowan-Robinson (1996)
and the evolutionary synthesis models
of Mazzei \& de Zotti (1996). Figure 10 demonstrates that in the absence
of sufficient observational data to constrain models between 
$800\mu \rightarrow 2\mu$,   
the adoption of a single average grain temperature (50K) 
provides extremely good agreement, to within 20\%, between 
the FIR and bolometric luminosities (and also the current SFRs) for
galaxies expected to be undergoing vigorous starformation. 

\begin{figure}
\begin{picture}(220,250)
\put(0,0){\includegraphics{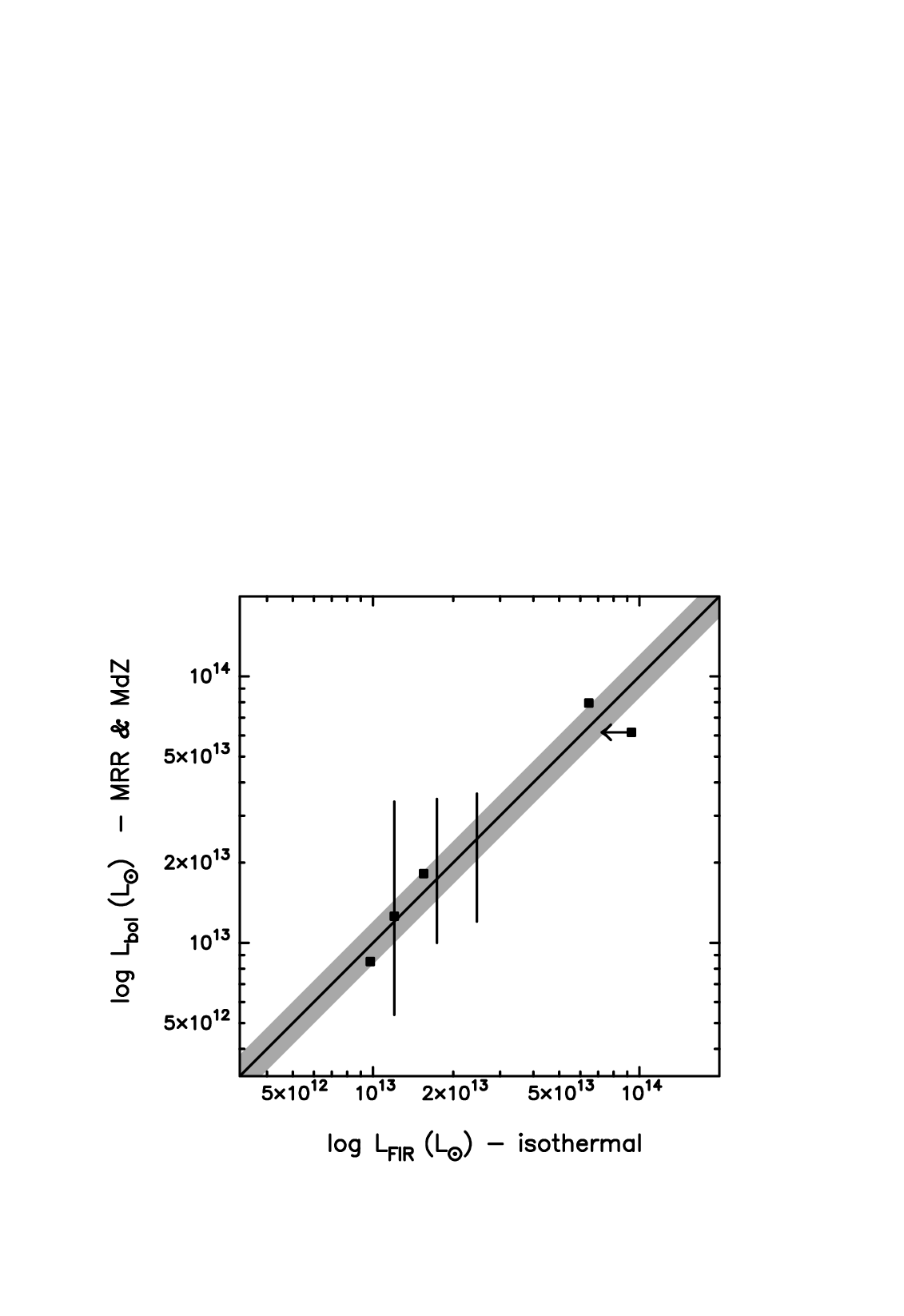}}
\end{picture}
\caption{
A comparison of the rest-frame FIR luminosities ($L_{FIR}$) 
calculated from a 50K 
isothermal model (table 3) and the bolometric luminosities ($L_{bol}$) determined
from radiative transfer models of Rowan-Robinson (1996, MRR) - filled
squares, and from the range of evolutionary synthesis models of Mazzei 
\& de Zotti (1996, MdZ) - lines, which provide a reasonable fit to the submm,
optical and IR data. The shaded region represents a
maximum discrepancy of 20\% about a perfect correlation between the FIR
luminosities, and hence SFRs, calculated from isothermal and more
sophisticated models.}
\end{figure}
     
\subsection{Molecular gas masses in high-redshift galaxies}

A measure of the total ($H\small{I} + H_{2}$)
gas mass of a high-redshift galaxy is a very useful
indicator of its evolutionary status because, as long as one can be
reasonably confident
that the galaxy in question is the progenitor of a present-day giant
elliptical (as is the case for radio galaxies), it provides a measure of
the fraction of the galaxy which has yet to be turned into stars at the
epoch of observation. One way to determine the gas mass of a galaxy is to
estimate its molecular ($H_2$) mass from a measurement of the intensity
of CO.
Unfortunately, in the absence of lensing,
it has proved impossible 
to detect the molecular gas in high-z galaxies directly 
(van Ojik {\it et al.} 1997, Evans {\it et al.} 1996, Barvainis \&
Antonucci 1996), with the possible exception of BR1202$-$0725 (Omont {\it
et al.} 1996, Ohta {\it et al.} 1996).
However, an alternative and potentially more productive approach is to
convert dust mass to gas mass by adopting a `reasonable' value for the 
gas:dust ratio, $M_{H_{2}}/M_{d}$.
However $M_{H_{2}}/M_{d}$ is not a well determined quantity in galaxies 
in the local universe, 
let alone at high redshift.
Studies of damped Lyman-$\alpha$ systems (DLAAS) currently provide the only 
opportunity to directly 
measure the dust content of the universe at early epochs. 
At $z \sim 3$ it has been suggested that the gas-dust ratio in DLAAS is
400--2000 (Fall, Pei \&
McMahon 1989, Pettini {\it et al.} 1994), a value significantly higher than 
the galactic value of 100--160 (Hildebrand 1983, Savage \& Mathis 1979), 
and on average higher than that found in nearby spirals ($\sim 500$, 
Devereux \& Young 1990),
ellipticals ($\sim 700$, Wilkind {\it et al.} 1995) and ULIRGS 
($540 \pm 290$, Sanders {\it et al.} 1991). 

While it may be true that the ratio $M_{H_{2}}/M_{d}$  
in high-redshift Lyman-$\alpha$ 
absorbers is significantly greater than in present day galaxies,
it seems likely that high-redshift radio galaxies, being the progenitors
of present day ellipticals, are considerably more evolved
than Lyman-$\alpha$ absorbers at comparable redshift (which, it has been 
argued, are the progenitors of present-day disc galaxies). Thus there
seems to be no clear justification for adopting an extreme gas:dust ratio
and so we have decided to adopt 
a conservative value of $M_{H_{2}}/M_{d} \sim 500$, consistent
with the values reported for galaxies in the local universe (the
appropriateness of this chosen value is discussed further in the next
section).

The resulting values of $M_{H_2}$ determined from continuum measurements
are listed in Table 3, and, are $\leq 8\times 10^{11} M_{\odot}$. These 
estimates of the $H_{2}$
gas mass limits are 
consistent with the failure to detect large reservoirs of
molecular gas directly in various CO transitions
(Evans {\it et al.} 1996; van Ojik {\it et al.} 1997, 
Barvainis \& Antonucci 1996) a fact which lends some circumstantial
support to our adopted gas:dust ratio.
Nevertheless, these gas masses are still extremely large and represent a
significant fraction of the present day stellar mass of the largest
elliptical galaxies. The crucial issue of whether this `significant
fraction' could be sufficiently large to indicate that these objects
deserve to be described as `prim\ae val' is considered below in the
final section of this paper.

\section{Conclusions: The Evolutionary Status of High-Redshift Galaxies}

We have presented sensitive continuum measurements at 
800$\mu m$ of a sample of
high-redshift  radio galaxies and quasars. These observations were motivated by
the goal of determining their evolutionary status in a relatively model
independent manner via a measure of their dust masses, star-formation
rates,
and molecular gas masses.
In addition to our 800$\mu m$ data presented here,
we have taken the opportunity to gather together all other published 
millimetre and submillimetre observations of high-redshift galaxies and 
quasars at $z > 2$, and Table 3, in effect,  
summarizes the current status of sub-millimetre cosmology, 

While it is expected that this subject will be revolutionized by 
the significantly more sensitive observations which should be possible
in the very near future using new submillimetre continuum bolometer
arrays such as SCUBA (Gear \& Cunningham 1995), we have argued in this
paper that there is much to be learned from considering the uncertainties
which afflict the reliable interpretation of the existing, albeit sparse
dataset. In this spirit we conclude by considering 
what can be deduced about the evolutionary status of high-redshift galaxies
from the numbers presented in Table\,3.

To focus the argument we adopt the strict (and deliberately extreme)
definition that 
a genuinely young elliptical galaxy would be expected to display a very high 
star-formation rate ($> 500 M_{\odot} yr^{-1}$), and to contain a total
$H\small{I}\,+\,H_{2}$ gas mass equivalent to the majority
of the stellar mass found in the most massive 
present-day giant ellipticals ({\it e.g.} $M_{H\small{I} + H_{2}}
\geq  10^{12} M_{\odot}$). The adoption of such a large present-day mass
can be justified on two levels. First, if one wants to unambiguously prove
that a given high-redshift object is prim\ae val on the basis of its gas
mass, one needs to demonstrate that, even if its eventual 
present-day stellar mass is extremely large, the bulk of it 
remains in gaseous form at the high-$z$ epoch of observation. Second,
since it is now well-established that the most powerful radio sources and
quasars in the low-redshift universe reside in host galaxies with
luminosities several times greater than $L^{\star}$ (Taylor {\it et al.}
1996), it is not unreasonable to assume that the considerably more
luminous active objects listed in Table 3 lie in still deeper potential
wells.

Since we are unable to detect the
contribution from 21\,cm HI emission, 
which is shifted to 250--500MHz at high-redshift, we use the mean ratio of 
M$_{HI}/M_{H_{2}} \sim 1.0 \pm 0.9$ found in elliptical, early-type and 
interacting  galaxies (Lees \etal\ 1991, Young \& Knezek 1989), to place
a lower mass limit of  M$_{H_{2}} \geq 5 \times 10^{11} M_{\odot}$ 
on the molecular gas content of the most luminous prim\ae val or 
proto-elliptical galaxies.

As illustrated in Figure\,9, 
when judged against these criteria of SFR and M$_{H_{2}}$, 
although several of the detected
objects in Table\,3 may perhaps have adequate star-formation rates, 
only one source, the quasar BR1202$-$0725, lies within the parameter 
space populated by prim\ae val galaxies under the above definition.
Given that both IRAS10214$+$4724 and the Cloverleaf quasar would also have
qualified as genuinely young on the basis of the H$_{2}$
gas mass prior to correcting for the estimated effects of lensing, the
suspicion, borne out by the recent CO detection (Omont \etal\ 1996, Ohta
\etal\ 1996)  must be that BR1202$-$0725 may also have amplified 
emission. Therefore based on our
analysis, the sub-millimetre properties of high-redshift AGN, which
have been identified at optical/IR/radio wavelengths, are more typical
of a strong starburst in an otherwise well-formed galaxy than of
genuinely prim\ae val objects. Such a result could indicate that we are
seeing the {\it final} stages of formation of these massive objects, or a
starburst triggered by a galaxy-galaxy interaction which is also
responsible for the observed AGN activity (in a heirarchical picture of
galaxy formation these two alternatives could 
essentially amount to the same thing).

In judging the robustness of the basic conclusion that these high-z
galaxies are {\it not} prim\ae val, it is important to consider in turn 
the effect of
the dust temperature and adopted gas:dust ratio on the dust masses 
($\rightarrow M_{H_{2}}$)  and FIR luminosities ($\rightarrow SFR$)
in Table\,3 since both of the latter quantities are
determined from an isothermal grey-body fit to the submillimetre continuum. 

First, the combined effect of the
temperature uncertainty (T$_{d} \sim 50 \pm 20$K)
on the FIR luminosities and dust masses,
which was discussed at length in \S3.3.2, 
is illustrated in the representative locus passing through 4C41.17 in 
figure\,9. 
This locus, which is appropriate for galaxies at $z = 2 - 5$ 
shows that the upper-limit in temperature (70K) moves the
high-z sources away from the parameter space occupied by prim\ae val
galaxies and into a region requiring extremely high star-forming
efficiencies ($\gg 100 L_{\odot}/M_{\odot}$). Conversely, while reducing the
temperature to 30K can increase the inferred gas mass to close to the
required value, this is at the expense of the FIR luminosity and 
star-formation rate 
which is reduced to $< 500 M_{\odot} yr^{-1}$.
In summary, our basic conclusion -- that we are {\it not} seeing the primary
formation event of these galaxies -- is, with the possible exception of
4C41.17, unaffected by 
varying the temperature of the dust between reasonable limits.

It is worth noting that while our 
adoption of a dust temperature of 50 K for these high-redshift objects results 
in values for $L_{FIR}/M_{H_{2}} \simeq 100$ somewhat greater than 
is seen in isolated
systems at low-redshift ($12 \pm 3 L_{\odot}/M_{\odot}$)
(Young \etal\ 1986, Solomon \& Sage 1988), such values are not unreasonable
and are comparable to that seen in the most luminous 
low-redshift interacting galaxies ($78 \pm 14 L_{\odot}/M_{\odot}$).
Indeed the location of the high-redshift sources  
on Figure 9
seems perfectly consistent with an extrapolation of the upper envelope of
star-forming efficiency defined by the most luminous ULIRGs.

Second, we address the issue of whether the uncertainty in the
assumed ratio $M_{H_{2}}/M_{d}\sim 500$ (\S 4.3) can affect our conclusions. 
Obviously it can if a `reasonable' range of $M_{H_{2}}/M_{d}$ at 
high-redshift is taken to extend to values of several 
thousand as suggested by the analysis of Fall \etal\ (1989). However, as
already discussed above, the relevance of the
investigations of Lyman $\alpha$ absorbers to the present study
of high-redshift AGN is dubious. 
More significant is the good agreement between our adopted value  and
the mean $M_{H_{2}}/M_{d}$ ratio, ($426 \pm 96$),
measured in the only 3 high-z sources (H1413$+$117, IRAS10214$+$4724, 
BR1202$-$0725) with confirmed CO line and submillimetre continuum detections.
The gas:dust ratio in these high-$z$ AGN should be relatively 
unaffected by lensing
assuming that, as in low-$z$ galaxies, the CO and submillimetre 
continuum have similar spatial distributions, and hence similar
amplifications. Thus, taken at face value, these few CO detections of 
high-redshift objects support our view that $M_{H_{2}}/M_{d} \sim 500$
is appropriate for luminous galaxies  out to $z \simeq 5$, 
and explains why
sub-millimetre continuum observations of high-redshift objects have been
more successful than attempts to detect the molecular 
gas directly through CO line
observations (and casts even more doubt on the claimed CO detection of
53W002, Yamada \etal\ 1995). 

To summarize, figure 9 demonstrates that adoption of a dust temperature of
50 K and a gas:dust ratio of 500 leads one to conclude that the
high-redshift objects listed in Table 9 would be better described as highly
efficient starburst galaxies than genuinely prim\ae val objects. This
conclusion is relatively immune to alteration of the adopted dust
temperature. Indeed, to alter it significantly 
requires the adoption of a gas:dust ratio $\simeq 5$ times
greater than has been assumed, and the low success rate
of molecular CO line observations argues against this option.
However, to finish on a cautionary note, there is one way to increase the
inferred gas mass without increasing its detectability through line
observations, and that is to relax the assumption of an
Einstein-de-Sitter Universe. Ignorance of $\Omega_0$ is of course a
problem which afflicts studies of galaxy evolution at all wavelengths
but, as discussed in section 3.4, at $z \simeq 4$ adoption of $\Omega_0
\simeq 0.1$ instead of $\Omega_0 = 1$ increases the inferred dust mass by a 
factor of 5 and so, all other things being equal, 
would increase the values of {\it both} $L_{FIR}$ and $M_{H_2}$ plotted
in Figure 9 by a factor of 5 {\it for the
high-redshift objects only}. Under these circumstances, 
with star-formation rates ($\simeq
10^3 M_{\odot} yr^{-1}$) and molecular gas masses ($M_{H_2} \simeq 5
\times 10^{11} M_{\odot}$) it would be hard to escape the conclusion
that essentially all of the high-redshift objects so far detected at
sub-millimetre wavelengths are in fact prim\ae val galaxies (one
could, for example, reduce the inferred gas mass by adopting a high dust
temperature such as $T \simeq 70$K, but this would in turn 
imply a truly prodigious star-formation rate).
 
Thus the unambiguous determination of the evolutionary status of the
high-redshift objects detected to date at sub-millimetre wavelengths must
await improved constraints on $\Omega_0$. However, we regard it as
encouraging for the future of sub-millimetre cosmology 
that the uncertainties peculiar to the interpretation of
these sub-millimetre data can be justified as being comparable in size to
the current cosmological uncertainties.  
In general, continuum observations at sub-millimetre and millimetre
wavelengths have proven to be more successful and have 
provided a comparable or lower H$_{2}$ mass limit than spectral line data 
obtained in an equal integration time. This observing efficiency,
together with an expected improvement in our understanding of the 
dust emissivity and gas-to-dust ratio at high-redshift, justifies the
continuation of this approach with the future generation of
sub-millimetre continuum instruments ({\it e.g.} SCUBA) which 
will be sensitive to SFRs $< 100$\,M$_{\odot}yr^{-1}$ 
at $z = 3 - 5$ and hence will
detect unlensed versions of IRAS10214$+$4724 and H1413$+$117 with ease and 
also detect the equivalent of low-redshift ULIRGS ({\it e.g.} Arp220, Mrk231)
at $z > 4$. Given that the sub-millimetre properties of the high-redshift 
radio galaxies and quasars studied to date point towards a more dramatic
formation event at yet higher redshift, it will be of particular interest 
to discover whether any of the objects detected by sub-millimetre surveys 
meet the criteria defined in Figure 9 for a genuinely prim\ae val giant
elliptical galaxy.

\vspace*{2.0in}

\section*{ACKNOWLEDGEMENTS}
We are indebted to the referee for helpful comments and suggestions that
led to the improvement of this paper. 
We thank M.S.Yun and N.Z. Scoville for communicating their unpublished 105\,GHz 
measurement of B20902+34. We are also grateful to
Jason Stevens and Rob Ivison for confirming our 800$\mu m$ observation
of H1413$+$117.
The James Clerk Maxwell Telescope is operated by The Observatories on 
behalf of the Particle Physics and Astronomy Research Council of the 
United Kingdom, the Netherlands Organisation for Scientific Research, 
and the National Research Council of Canada. We thank Jeff Cox, Alan
Hatakeyama, Ed Lundin, Rusty Luthe, Kimberley Pisciotta and Jim Pomeroy for their 
valuable assistance at the telescope.
DHH gratefully acknowledges receipt of a PPARC PDRA during the course of
this work. 

\section*{REFERENCES}

\frenchspacing
Andreani P., La Franca F., Cristiani S., 1993, MNRAS, 261, L35\\
Barvainis R., Antonucci R.R.J.,  1996, PASP, 108,187 \\
Barvainis R., Tacconi L., Antonucci R.R.J., Alloin D., Coleman P.,
1994, Nat. 371, 586 \\
Barvainis R., Antonucci R.R.J., Colelman P., 1992, ApJ, 399, L19\\
Barvainis R., Antonucci R.R.J., Hurt T., Colelman P., Reuter H.-P.,
1995, ApJ, 451, L9\\
Bazell D., D\'{e}sert F.X., 1988, ApJ, 333, 353 \\
Bower R.G., Lucey J.R., Ellis R.S., 1992, MNRAS, 254, 589 \\
Carilli C.L., Owen F.N., Harris D.E., 1994, AJ, 107, 480 \\
Chambers K.C., Charlot S., 1990, ApJ, 348, L1\\
Chambers K.C. \& McCarthy P.J., 1990, ApJ, 354, L9\\
Chambers K.C., Miley G.K., van Breugel W., 1987, Nat, 329, 604\\
Chambers K.C., Miley G.K., van Breugel W., 1990, ApJ, 363, 21\\
Chini R., Kr\"{u}gel E., Kreysa E., 1986, A\&A, 167, 315\\
Chini R., Kreysa E., Biermann P., 1989a, A\& A, 219, 87 \\
Chini R., Kr\"{u}gel E., Kreysa E., Gem\"{u}nd H.-P., 1989b, A\&A, 216, L5\\
Chini R., Kr\"{u}gel E., 1994, A\&A, 288, L33 \\
Cimatti A., Freudling W., 1995, A\&A, 300, 366 \\
Diamond P.J., Goss W.M., Romney J.D., Booth R.S., Kalberla P.M.W.,
Mebold U., 1989, ApJ, 347, 302 \\
Devereux N.A., Young J.S., 1990, ApJ, 359, 42 \\   
Dickinson M., 1997, In: {\it HST and the High-Redshift Universe}, 
eds. N.Tanvir, A.Aragon-Salamanca, J.Wall, World Scientific Press, in
press. \\ 
Downes D., Solomon P.M., Sanders D.B., Evans A.S., 1996, A\&A, 313, 91 \\ 
Draine B.T., 1990, In: {\it The Interstellar Medium in Galaxies}, p.483,
eds. H.A. Thronson and J.M. Shull, Kluwer, Dordrecht\\
Draine B.T., Lee H.M., 1984, ApJ, 285, 89\\
Duncan W.D., Robson E.I., Ade P.A.R., 
Griffen M.J., Sandell G., 1990, MNRAS, 243,126\\
Dunlop J.S., 1997, In: {\it HST and the High-Redshift Universe}, 
eds. N.Tanvir, A.Aragon-Salamanca, J.Wall, World Scientific Press, in
press. \\ 
Dunlop J.S., Peacock J.A., 1993, MNRAS, 263, 936\\
Dunlop J.S., Hughes D.H., Rawlings S., Eales S.A., Ward M.J, 1994,
Nat., 370, 347 \\ 
Dunlop J.S., Peacock J., Spinrad H., Dey A., Jimenez R., Stern D.,
Windhorst R., 1996, Nat., 381, 581 \\
Eales S.A., Rawlings S., 1993, ApJ, 411, 67\\
Eales S.A., Rawlings S., 1996, ApJ, 460, 68\\
Eales S.A., Rawlings S., Puxley P., Rocca-Volmerange B., Kuntz K., 
1993, Nat, 363, 140\\
Eisenhardt P., Dickinson M., 1992, ApJ, 399, L47\\
Eisenhardt P., Armus L., Hogg D.W., Soifer B.T., Neugebauer G., Werner
M.W., 1996, ApJ, 461, 72 \\
Evans A.S., Sanders D.B., Mazzarella J.M., Solomon P.M., Downes D.,
Kramer C., Radford S.J.E., 1996, ApJ, 457, 658 \\
Fall M.S., Pei Y.C., McMahon R.G., 1989, ApJ, 341, L5\\
Fukugita, Hogan, Peebles 1996, Nat., 381, 489\\
Gautier T.N., Boulanger F., Perault M., Puget J.L., 1992, AJ, 103, 1313 \\
Gear W.K., Cunningham C.,  1995, In: {\it Multifeed systems for radio 
telescopes}, P.A.S.P. Conf Ser., Vol. 75, p.215, eds. D.T.Emerson, 
J.M.Payne \\  
Giavalisco M., Steidel C.C., Macchetto F.D., 1996, ApJ, 470, 189 \\
Guiderdoni B., Rocca-Volmerange B., 1987, A\&A, 186, 1\\
Heckman T.M., Chambers K.C., Postman M., 1994, ApJ, 391, 39\\
Helou G., Beichman C.A., 1991, In: {\it From Ground-Based to Space-Borne
Sub-mm Astronomy},  p.117, ESA SP-314 \\
Hildebrand R.H., 1983, QJRAS, 24, 267 \\
Hughes D.H., 1996, In: {\it Cold Gas at High Redshift}, eds. M.N.Bremer,
P.P.van der Werf, H.J.A.Rottgering, C.L.Carilli, Kluwer, p.311 \\   
Hughes D.H., Appleton P.N., Schombert, J.M. 1991, ApJ, 370, 176 \\
Hughes D.H., Davies R., Ward M.J., 1997, in preparation\\
Hughes D.H., Gear W.K., Robson E.I., 1994, MNRAS, 270, 641 \\
Hughes D.H., Robson E.I., Dunlop J.S., Gear W.K., 1993, MNRAS, 263, 607\\
Illingworth G., 1997, In: {\it HST and the High-Redshift Universe}, 
eds. N.Tanvir, A.Aragon-Salamanca, J.Wall, World Scientific Press, in
press. \\ 
Isaak K., McMahon R., Hills R., Withington, S., 1994, 269, L28 \\
Ivison R.J., 1995, MNRAS, 275, L33 \\
Kennicutt R.C., 1983, ApJ, 272, 54 \\
Knapp G.R., Patten B.M., 1991, AJ, 101, 1609\\
Lacy \etal\ 1994, MNRAS, 271, 504 \\
Laing, Riley \& Longair 1983, MNRAS, 204, 151 \\ 
Lees J.F., Knapp G.R., Rupen M.P., Phillips T.G., 1991, ApJ, 379, 177\\
Lilly S.J., 1988, ApJ, 333, 161\\
Lilly S.J., 1989, ApJ, 340, 77\\
Lockman F.J., Johoda K., McCammon D., 1996, ApJ, 302, 432 \\
Low F.J. \etal\ 1984, ApJ, 278, L19  \\
Magain P., Surdej J., Swings J., Borgeest U., Kayser R., 1988, Nature, 
334, 325 \\
Mathis J.S., Whiffen G., 1989, ApJ, 341, 808 \\ 
Mazzarella J.M., Graham J.R., Sanders D.B., Djorgovski S., 1993, ApJ,
409, 170\\
Mazzei P., de Zotti G., 1996, 279, 555 \\  
Mazzei P., de Zotti G., Xu C., 1994, ApJ, 422, 81  \\  
McCarthy P.J., Spinrad H., van Breugel W., 1995, ApJS, 99, 27\\
McCarthy P.J., Spinrad H., Djorgovski S., Strauss M.A., van Breugel W., Liebert J., 1987a, ApJ, 319, L39 \\
McCarthy P.J., van Breugel W., Spinrad H., Djorgovski S., 1987b, ApJ, 321, L29 \\
McMahon R.G., Omont A., Bergeron J., Kreysa E., Haslam C.G.T., 1994, MNRAS, 267, L9 \\ 
Meyer D.M., 1990, ApJ, 364, L5 \\
Miley G.K., Chambers K.C., van Breugel W., Macchetto F., 1992, ApJ, 401, L69\\
Miller G.E.,  Scalo J.M., 1979, ApJS, 41, 513 \\
Ohta K., Yamada T., Nakanishi K., Kohno K., Akiyama M., Kawabe R., 1996, Nature, 382, 426 \\
Omont A., Petitjean P., Guilloteau S., McMahon R.G., Solomon P.M., Pecontal E., 1996, Nature, 382, 428 \\
Omont A., McMahon R.G., Cox P., Kreysa E., Bergeron J., Pajot F.,
Storrie-Lombardi,L.J., 1996, A\&A, 315, 1 \\ 
Owen F.N., Laing R.A., 1989, MNRAS, 238, 357 \\ 
Pettini M., Smith L.J., Hunstead R.W., King D.L., 1994, ApJ, 426, 79 \\ 
Rawlings S., Saunders R.D.E., 1991, Nat, 349, 138\\
Rawlings S., Lacy M.,Blundell K.M., Eales S.A., Bunker A.J., Garrington
S.T.,1996, Nature, 383, 502 \\
Rowan-Robinson M., 1986, MNRAS, 219, 737\\
Rowan-Robinson M., 1992, MNRAS, 258, 787\\
Rowan-Robinson M., Efstathiou A., Lawrence A., Oliver S., 
Taylor A., Broadhurst T.J., McMahon R.G., Benn C.R., Condon J.J.,
Lonsdale C.J., Hacking P., Conrow T., Saunders W.S., Clements D.L., Ellis R.S.,
Robson I., 1993, MNRAS, 261, 513 \\
Sandage A. 1972, ApJ, 271, 21 \\
Sanders D.B., Soifer B.T., Elias J.H., Madore B.F., Matthews K., 
Neugebauer G., Scoville N.Z., 1988, ApJ, 325, 74 \\
Sanders D.B., Scoville N.Z, Soifer B.T., 1991, ApJ, 370, 158 \\
Sandell G., 1994, MNRAS, 271, 75 \\
Savage B.D., Mathis J.S., 1979, Ann. Rev. Astr. Ap., 17, 73 \\
Schmidt M., Green R.F., 1983, ApJ, 269, 352 \\
Schneider D.P., Schmidt M., Gunn J.E., 1991, AJ, 101, 2004 \\
Scoville N.Z., Young J.S., 1983, ApJ, 265, 148\\
Solomon P.M., Sage L, 1988, ApJ, 334, 613 \\ 
Spinrad H., Dey A., Graham J.R., 1995, ApJ, 438, L51 \\
Steidel C.C., Giavalisco M., Pettini M., Dickinson M., Adelberger K.L., 
1996,  ApJ, 462, L17 \\
Stockton A., Kellogg M., Ridgway S.E., 1995, ApJ, 443, 69 \\
Tadhunter C.N., Scarrott S.M., Draper P., Rolph C., 1992, MNRAS, 256, 53p\\
Taylor G.L., Dunlop J.S., Hughes D.H., Robson E.I., 1996, MNRAS, 283,
930 \\
Terebey S., Fich M., 1986, ApJ, 309, L79 \\
Thronson H., Telesco C., 1986, ApJ, 311, 98\\ 
van Ojik R., R\"{o}ttgering H.J.A., van der Werf P.P., Miley G.K.,
Carilli C.L., Isaac K., Lacy M., Jenness T., Sleath J., Visser AA., Wink
J., 1997, A\&A, in press \\
van Steenberg M.E., Schull J.M., 1988, ApJ, 335, 197\\  
Windhorst R.A., Burstein D., Mathis D.F., Neuschaefer L.W., 
Bertola F., Buson L.M., Koo D.C., Matthews K., Barthel P.D., Chambers K.C.,
1991, ApJ, 380, 362\\
Windhorst R.A., Gordon J.M., Pascarelle S.M., Schmidtke P.C., Keel W.C.,
Burkey J.M., Dunlop J.S., 1994, ApJ, 435, 577 \\ 
Wang B., 1991, ApJ, 374, 456 \\
Wilkind T., Henkel C., 1995, A\&A, 297, L71 \\
Yamada T., Ohta K., Tomita A., Takata T., 1995, AJ, 110, 1564 \\
Young J.S., Knezek P.M., 1989, ApJ, 347, L55 \\ 
Young J.S., Kenny J.D., Tacconi L., Claussen M.J., Huang Y.-L.,
Tacconi-Garman L., Xie S., Schloerb F.P,  1986, ApJ, 311, L17 \\
Yun M.S., Scoville N.Z., 1996, private communication \\ 
Zepf S.E., Silk J., 1996, ApJ, 466, 114 \\
\end{document}